\documentclass[a4paper,onecolumn,11pt,accepted=2023-04-26]{quantumarticle}

\pdfoutput=1

\usepackage{amsfonts}
\usepackage{amssymb}
\usepackage{amsmath}
\usepackage{bbm}
\usepackage{graphicx}
\usepackage{verbatim}
\usepackage{hyperref}
\usepackage[english]{babel}
\usepackage[numbers]{natbib}
\usepackage[pdftex]{color}
\usepackage{dsfont}
\usepackage{wrapfig}
\usepackage[normalem]{ulem}
\usepackage{comment}
\usepackage[dvipsnames]{xcolor}
\usepackage{graphicx}
\usepackage{alltt}

\hypersetup{
    colorlinks=true,       
    linkcolor=cyan,         
    citecolor=magenta,        
    filecolor=magenta,      
    urlcolor=cyan,          
    runcolor=cyan
}

\usepackage{lipsum}
\usepackage{mdframed}
\usepackage{booktabs}
\usepackage{longtable}

\usepackage{natbib}
\setcitestyle{numbers,sort&compress}
\usepackage{graphicx}
\usepackage{svg}
\usepackage{bm}
\usepackage[caption=false]{subfig}
\usepackage{color}
\usepackage{hyperref}  
\usepackage{soul}
\usepackage{physics}
\usepackage{bbm}
\usepackage[ruled, lined, linesnumbered, commentsnumbered, longend]{algorithm2e}

\usepackage{tikz}
\usetikzlibrary{quantikz}

\newcommand{\beq}{\begin{eqarray}}
\newcommand{\eeq}{\end{eqarray}}
\newcommand{\bes} {\begin{subequations}}
\newcommand{\ees} {\end{subequations}}
\newcommand{\hh}{\mathcal{H}}

\newcommand{\bb}{\mathcal{B}}

\newtheorem{theorem}{Theorem}

\newtheorem{definition}{Definition}
\newtheorem{lemma}{Lemma}
\newtheorem{example}{Example}

\renewcommand{\Re}{\operatorname{Re}}
\renewcommand{\Im}{\operatorname{Im}}

\newcommand{\ignore}[1]{}

\begin{document}

\title{Two-Unitary Decomposition Algorithm and Open Quantum System Simulation} 

\author{Nishchay Suri}
\affiliation{QuAIL, NASA Ames Research Center, Moffett Field, California 94035, USA}
\affiliation{USRA Research Institute for Advanced Computer Science, Mountain View, California 94043, USA}
\affiliation{Department of Physics, Carnegie Mellon University, Pittsburgh, Pennsylvania 15213, USA}
\email{nsuri@usra.edu}

\author{Joseph Barreto}
\affiliation{QuAIL, NASA Ames Research Center, Moffett Field, California 94035, USA}
\affiliation{USRA Research Institute for Advanced Computer Science, Mountain View, California 94043, USA}
\affiliation{QuTech, Delft University of Technology, Delft, The Netherlands}
\email{joeybarreto4@gmail.com}

\author{Stuart Hadfield}
\affiliation{QuAIL, NASA Ames Research Center, Moffett Field, California 94035, USA}
\affiliation{USRA Research Institute for Advanced Computer Science, Mountain View, California 94043, USA}

\author{Nathan Wiebe}
\affiliation{Department of Computer Science, University of Toronto, Toronto, Ontario M5S 3E1, Canada}
\affiliation{Pacific Northwest National Laboratory, Richland, Washington 99352, USA}

\author{Filip Wudarski}
\affiliation{QuAIL, NASA Ames Research Center, Moffett Field, California 94035, USA}
\affiliation{USRA Research Institute for Advanced Computer Science, Mountain View, California 94043, USA}
\email{filip.a.wudarski@nasa.gov}

\author{Jeffrey Marshall}
\affiliation{QuAIL, NASA Ames Research Center, Moffett Field, California 94035, USA}
\affiliation{USRA Research Institute for Advanced Computer Science, Mountain View, California 94043, USA}
\email{jeffrey.s.marshall@nasa.gov}

\begin{abstract}
Simulating general quantum processes that describe realistic interactions of quantum systems following a non-unitary evolution is challenging for conventional quantum computers that directly implement unitary gates.
We analyze complexities for promising methods such as the Sz.-Nagy dilation and linear combination of unitaries that can simulate open systems by the probabilistic realization of non-unitary operators, requiring multiple calls to both the encoding and state preparation oracles. We propose a quantum two-unitary decomposition (TUD) algorithm to decompose a $d$-dimensional operator $A$ with non-zero singular values as $A=(U_1+U_2)/2$ using the quantum singular value transformation algorithm, avoiding classically expensive singular value decomposition (SVD) with an $\mathcal{O}(d^3)$  overhead in time. 
The two unitaries can be deterministically implemented, thus requiring only a single call to the state preparation oracle for each.
The calls to the encoding oracle can also be reduced significantly at the expense of an acceptable error in measurements. 
Since the TUD method can be used to implement non-unitary operators as only two unitaries, it also has potential applications in linear algebra and quantum machine learning.
\end{abstract}
\maketitle

\section{Introduction}

In the early 1980s Manin \cite{manin1980computable} and Feynman \cite{feynman1982simulating} independently proposed that in order to circumvent the prohibitive scaling in the simulation of quantum systems, one would need a device that operates according to the principles of quantum mechanics. These considerations gave birth to the field that we now call {\it quantum computing} \cite{nielsen2002quantum}. Since then an entire zoo of quantum algorithms for simulating quantum processes has emerged, some of them promise spectacular improvements over their classical alternatives, even exponential speedups~\cite{nielsen2002quantum,lloyd1996universal,martyn2021,RevModPhys.86.153,montanaro2016quantum,preskill2021quantum}. However, most of these prominent quantum algorithms are solely tailored for unitary simulations of isolated quantum phenomena. Since such a quantum device is universal in its computational nature, one may expect that its applicability reaches beyond the standard unitary framework.

In general, quantum systems of interest are rarely isolated as they are exposed to the inevitable influence of their environments.
The environment consists of inaccessible degrees of freedom of the system's natural surroundings as well as from the control devices that can become entangled with the system, ultimately leading to a loss of information. 
Therefore, the dynamics of a general quantum process must be described by a non-unitary, open quantum evolution~\cite{nielsen2002quantum,breuer2002theory,lindblad1976generators,gorini1976completely}. Understanding open quantum dynamics is indispensable to the study of dissipation and decoherence in quantum systems and therefore, noise in quantum circuits~\cite{gyongyosi2018survey,RevModPhys.86.1203,PhysRevLett.82.2417,RevModPhys.88.041001,PhysRevA.87.012324,dissi-lindblad-zanardi}, and is fundamental to a wide variety of phenomena such as thermalization~\cite{PhysRevE.81.051135,kastoryano2016quantum}, non-equilibrium steady states~\cite{PhysRevA.88.043635,prosen2009matrix,PhysRevLett.106.217206}, transport in strongly correlated systems~\cite{PhysRevB.80.035110,PhysRevB.86.125118} as well as applications in quantum biology~\cite{huelga2013vibrations,hu2021general,mostame2012quantum,PhysRevLett.108.020602}. Moreover, open quantum dynamics can be harnessed to perform universal quantum computation~\cite{verstraete2009quantum, coherent-ss-zanardi}  and prepare topological~\cite{PhysRevA.91.042117,diehl2011topology,bardyn2013topology} as well as entangled states~\cite{PhysRevA.78.042307,reiter2016scalable,kastoryano2011dissipative, q-data}. Simulating the dynamics of open quantum systems is vital to understanding these processes.

Unlike isolated quantum evolution, open quantum evolution is non-unitary and thus presents a challenge for conventional quantum computers that are capable of directly implementing unitary gates. Quantum algorithms such as trotterization of Lindbladians~\cite{kliesch2011dissipative,wang2011quantum,barthel2012quasilocality,PhysRevLett.127.020504} and simulation of Markovian open dynamics from a set of universal channels~\cite{bacon2001universal,sweke2015universal} rely on dilation techniques of encoding a non-unitary operator in a larger unitary. Recent advances have lead to promising methods such as the Sz.-Nagy encoding, which provides a minimal dilation~\cite{hu2020quantum,hu2021general,gaikwad2022simulating,head2021capturing}. Another useful method for realizing a non-unitary operator is by implementing it as a linear combination of unitaries (LCU)~\cite{10.5555/2481569.2481570,PhysRevLett.114.090502}, recently used to simulate Lindblad evolution~\cite{cleve_et_al:LIPIcs:2017:7477}. In Ref.~\cite{PhysRevLett.127.270503}, the authors propose implementing an arbitrary operator as a linear combination of four unitaries created by approximate operator exponentiation. A recent work~\cite{schlimgen2022quantum} implements the non-unitary operators in the singular basis by calculating the singular value decomposition (SVD) classically, thus incurring significant overhead for large systems. Additionally, non-unitary operators may be implemented via a combination of unitary operations and measurements requiring a feedback loop~\cite{lloyd_viola_2001,jiang_2017}, imaginary time evolution algorithms~\cite{motta2020determining,nishi2021implementation,PRXQuantum.2.010317} that need quantum tomography at each step, and even with methods that use the intrinsic dissipation of the quantum computer~\cite{sun2021efficient}.

In this article, we first analyze prominent techniques such as the Stinespring dilation, the Sz.-Nagy dilation and the LCU method and discuss their implementation of non-unitary operators to simulate open quantum dynamics and calculate expectation values of observables. 
The Sz.-Nagy and LCU methods implement the dynamical map by the probabilistic realization of each non-unitary Kraus operator individually. We estimate the number of encoding
oracle calls, state preparation oracle calls, and auxiliary qubits required for successful implementation of individual operators in the dynamical map. 
We provide a fully quantum method to estimate the expectation value of an observable to a specified confidence level, improving on previous methods that use Cholesky decomposition, which incurs $\mathcal{O}(d^3)$ of classical overhead, where $d$ is the dimension.
In particular, given an initial state $\ket{\psi}$ prepared by the state preparation oracle $\mathcal{S}$ and a contraction
operator $A$ block encoded as an isometry (unitary)~$U_A$,
the state $A\ket{\psi}/\sqrt{p}$ where $p=\bra{\psi}A^\dagger A \ket{\psi}$, $0<p\leq 1$, can be successfully implemented with probability at least $1-\beta$
using $\mathcal{O}(1/p \log(1/\beta))$ calls to both $U_A$ and $\mathcal{S}$. 
For realistic cases of interest such as the amplitude damping channel, a particular Kraus operator can have $p\ll 1$ which leads to a large number of calls to both $U_A$ and $\mathcal{S}$, which can be especially limiting if the initial state is difficult to prepare.

We provide a quantum algorithm to decompose any arbitrary $d$-dimensional contraction operator $A$ $(\| A\|\leq 1)$\footnote{Throughout $\|A\|$ denotes the operator norm of $A$, or equivalently, the maximum singular value of $A$.}, with non-zero singular values into two unitaries such that $A = (U_1+U_2)/2$ using the quantum singular value transformation (QSVT) algorithm~\cite{gilyen2019quantum,low2019hamiltonian,low2017optimal,martyn2021}, {without} explicitly computing the SVD, which has $\mathcal{O}(d^3)$ of classical overhead. Given a state $\ket{\psi}$ prepared by $\mathcal{S}$ and a block encoding $U_A$ of an arbitrary operator $A$, our two-unitary decomposition (TUD) algorithm implements $\tilde U_1\ket \psi,\tilde U_2\ket \psi$ which are $\epsilon$ approximations of $U_1\ket{\psi},U_2\ket{\psi}$ with probability at least $1-\beta$ using $\mathcal{O}(\log(1/\beta))$ calls to $\mathcal{S}$, and $\mathcal{O}(1/\delta \log(1/\epsilon) \log(1/\beta))$ calls to $U_A$ and $U_A^\dagger$, where the singular values $\sigma_i $ of $A$ are such that $ \sigma_i \in [\delta,1-\delta] \ \forall i$. Therefore, instead of implementing $A\ket{\psi}$ directly which is probabilistic, the TUD algorithm implements $\tilde U_1 \ket{\psi}$ and $\tilde U_2 \ket{\psi}$ deterministically with each requiring approximately one use of state preparation oracle $\mathcal{S}$ or a single shot of $\ket{\psi}$. 
This is a considerable improvement from the previously needed $\mathcal{O}(1/p \log(1/\beta))$ calls to $\mathcal{S}$. We show that the unitary decomposition can be used to estimate the expectation value of an observable with respect to $A\ket{\psi}$ which can be done with Hadamard tests. 
Since the TUD method can be used to implement non-unitary operators as only two unitaries, it also has potential applications in linear algebra and quantum machine learning~\cite{HHL2009,schuld2015introduction,biamonte2017quantum}.

\bigskip

\tableofcontents

\section{Open Quantum Systems and Kraus Operators \label{sec:OQS_overview}}

The evolution of an isolated system in a mixed quantum state is described by the von Neumann equation 
\begin{equation}\label{eq:vonNeumann}
    \dot{\rho}_t = -i[H(t),\rho_t] \;,
\end{equation}
where $H(t)$ is the system's Hamiltonian and $\rho_t:=\rho(t)$ is the density matrix (in $\hbar=1$ units).  The formal solution to Eq.~\eqref{eq:vonNeumann} is a unitary transformation
\begin{equation}
    \rho_t = \mathcal{U}_t(\rho_0)=U_t\rho_0 U_t^\dag,
\end{equation}
where $U_t =\mathcal{T}\exp(-i \int_0^t H(t^\prime)d t^\prime)$ and $\mathcal{T}$ is the time-ordering operator,  simplifying to $U_t = \exp(-iHt)$ when $H$ is time-independent.

The system and environment together undergo the unitary evolution of an isolated quantum process. To describe the effective dynamics of the system, we trace out the environmental degrees of freedom obtaining the open quantum evolution.
Without loss of generality, one can assume that the system is initially decoupled from its environment such that the entire density matrix is 
$\rho_0 = \omega_E\otimes \rho_S$, where  $\omega_E,\; \rho_S$ are the initial density matrices of the environment and system, respectively. The total evolution 
$U_t=\mathcal{T}\exp(-i \int_0^t H_{\textrm{Total}}(t') dt' )$ 
is described by a Hamiltonian of the general form $H_{\text{Total}}(t)=I_E\otimes H_S(t) +H_E(t) \otimes I_S  + H_I(t)$, where $H_S$, $H_E$ act only on the system and the environment respectively, while $H_I$ is responsible for their interaction. This allows us to describe the effective dynamics of the $d$-dimensional system in terms of a dynamical map $\Lambda_t\colon \bb(\mathcal{H}_S)\rightarrow \bb(\mathcal{H}_S)$ given by
\begin{equation}
    \Lambda_t(\rho_S) = \Tr_E\Big( U_t \; \omega_E \otimes \rho_S \; U_t^\dag\Big) = \sum_{k=1}^m A_k(t) \rho_S A_k^\dag(t) \;,
    \label{eq:kraus_derive}
\end{equation}
where $\mathcal{H}_S$ is the
Hilbert space of the system, $\Tr_E$ is the partial trace over the Hilbert space of the environment, and $m\leq d^2$.  
The dynamical map $\Lambda_t$ is a completely positive trace preserving (CPTP) map which describes physically valid transformation from quantum states to quantum states~\cite{breuer2002theory}. 
Following the partial trace, one obtains a set of Kraus operators $ A_k \in \mathcal B (\mathcal{H}_S)$, 
where $\mathcal B (\mathcal H_S)$ is the space of bounded linear operators. The Kraus operators are in general time-dependent, non-unitary, and also depend on the initial environmental state.  For example, if $\omega_E = \ket{\nu}\bra{\nu}$, then $A_k(t) \equiv (\bra{k} \otimes \mathbbm{1}_S) U_t (\ket{\nu} \otimes \mathbbm{1}_S)$, where $\ket{k}$ is a basis for the Hilbert space of the environment.  Furthermore, the Kraus representation of the dynamical map is not unique as the partial trace over the environment is independent of the choice of basis.  The Kraus operators themselves lack any special structure apart from the following trace-preservation constraint: $\sum_k A_k(t)^\dagger A_k(t) = I$ for all $t\ge0$. For notational clarity we shall suppress any explicit dependence on time or other parameters unless stated otherwise: $A_k(t) \equiv A_k$.

Since a quantum computer only has access to unitary operations and measurements, the implementation of dynamical maps comprised of non-unitary Kraus operators is a challenge. Implementing the transformation in Eq.~\eqref{eq:kraus_derive} by implementing the entire unitary~$U_t$ that describes the system and its environment together quickly becomes infeasible as the environment can often have large or even infinite dimensionality. Therefore it is necessary to develop quantum algorithms that can emulate the effect of environmental influence indirectly. 
Given Kraus operators $A_k$ block encoded in unitaries $U_k$, we estimate the number of resources to implement each non-unitary Kraus operator individually on a quantum computer as well as the entire map given by Eq.~\eqref{eq:kraus_derive}. 
In many practical cases, $m$ can be assumed to be $ \mathcal{O}(\text{poly}\log d)$ and therefore, the implementation of the map becomes practically feasible. For brevity we restrict further analysis to $\rho_S = \ket{\psi}\bra{\psi}$, where $\ket\psi$ is created by the oracle $\mathcal S$, with the extension to mixed states being straightforward. 
Additionally, we analyze the complexity to estimate the expectation value of a Hermitian observable $O$, 
\begin{align}
 \langle O \rangle = {\Tr\Big(O\Lambda(\ket \psi \bra \psi) \Big)} =  \sum_{k=1}^m  \langle \psi | A_k^\dag O A_k |\psi \rangle\;,
 \label{eq:Kraus_expect}
\end{align}
such that the estimator is $\widehat{\langle O \rangle}:={1}/{N}\sum_{i=1}^N x_i$ where $x_i$ is the $i^{\text {th}}$  measured (randomly sampled) value of the random variable $X$, where $\mathbb{E}(X)=\langle O\rangle$ 
and $N$ is the total number of measurements. In practice $N$ will be selected so that the variance of the estimate (or standard error of the mean) is below a certain pre-chosen threshold $Var(\widehat{\langle O\rangle}) = Var(X)/N \leq v$, as discussed in more detail in App.~\ref{app:MeasureExpect}.

\section{Analysis for Previous Methods}
\label{sec:Previous}

\subsection{Stinespring Dilation}
Simulating the dynamical map of Eq.~\eqref{eq:kraus_derive} by realizing the entire evolution of the system and its 
environment together as a unitary is 
often infeasible. A 
natural and intuitive 
first approach is to instead emulate the degrees of freedom of the environment by a set of auxiliary qubits that may be traced out to obtain the desired map.  The Stinespring dilation theorem, proposed long before the advent of quantum simulation, offers a way to achieve this `mimicking' of the environmental degrees of freedom.

\begin{theorem}[Stinespring Dilation theorem]
Let $\Lambda \colon \bb(\mathcal{H}) \rightarrow \bb(\mathcal{H})$ be a quantum CPTP map over a finite dimensional Hilbert space $\mathcal{H}$. Then there exists a Hilbert space $\mathcal{E}$, a unitary operator {$U_{\text{St}}$} over the joint Hilbert space $\mathcal{E} \otimes \mathcal{H} $, and a quantum state $\omega \in \bb(\mathcal{E})$, such that
\begin{equation}
    \Lambda (\rho) = \Tr_{{E}}\left( U_{\mathrm{St}} \; \omega \otimes \rho \; U_{\mathrm{St}}^\dag \right),\,\, \forall \rho \in \mathcal{H},
\end{equation}
where $\mathrm{dim}(\mathcal{E}) \le \mathrm{dim}(\mathcal{H})^2$, and the representation is unique up to a unitary equivalence. 
\end{theorem}

\begin{figure}[h!]
\centering
\begin{tikzpicture}
    \node[scale=1.] {
   		\begin{quantikz}[row sep={0.6cm,between origins}]
            \lstick[wires=2]{$\ket{0}^{\otimes \lceil\log m\rceil}$}  & \qw &\gate[wires=5,nwires=3][2cm]{U_{\text{St}}} & \qw & \trash{ } \\
             & \qw & & \qw & \trash{ } \\
             & & & & & \\
            \lstick[wires=2]{$\ket{\psi}$} & \qw & \qw & \qw & \qw \rstick[wires=2]{$\Lambda({\ket{ \psi} \bra { \psi}})$} \\
             & \qw & & \qw & \qw
        \end{quantikz}};
\end{tikzpicture}
\caption{The Stinespring dilation unitary $ U_{\text{St}}$ with $\lceil \log m\rceil$
auxiliary qubits that are traced out (indicated by downward-pointing arrows) to obtain the Kraus map.}
\label{fig:Stinespring}
\end{figure}
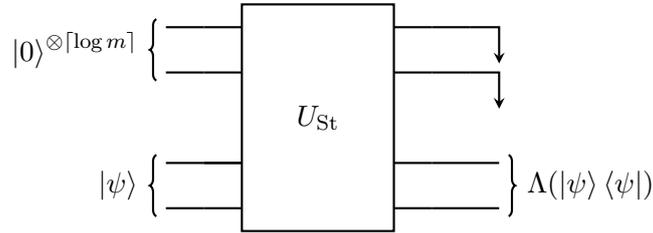

The Kraus operators can be embedded in the first block-column of the Stinespring dilation unitary $U_{\text{St}}$ that acts on the system and a set of auxiliary qubits emulating the environment
\begin{equation}\label{eq:stinespring_unitary}
    U_{\text{St}} = \sum_{k=1}^m |k\rangle\langle 0|\otimes A_k + \dots =     \begin{bmatrix} 
        A_1 & \cdots & \cdots \\ 
        A_2 & \cdots & \cdots \\
        \vdots & \ddots &\vdots\\
        A_m &\ldots & \ldots 
    \end{bmatrix}\;,
\end{equation} 
where $m\leq d^2$. We apply the unitary 
\begin{align}\label{eq:stinespring-action}
    U_{\text{St}} \ket{0}\otimes \ket{\psi}= \sum_{k=1}^m \ket{k}\otimes  A_k \ket{\psi} \;,
\end{align}
and trace out the auxiliary qubits--analogous to tracing out the environment--to obtain the dynamical map $\Lambda(\ket{\psi}\bra{\psi})=\sum_k A_k \ket{\psi}\bra{\psi} A^\dagger_k$.
Given the Kraus operators $A_k$, the $md$ dimensional matrix $U_{\text{St}}$ may be created by filling in the rest of $(m-1)d$ columns via the Gram-Schmidt process, incurring a classical overhead of $\mathcal{O}(m^3d^3)$.
The quantum circuit for realizing the unitary $U_{\text{St}}$ in Fig.~\ref{fig:Stinespring} uses $\mathcal{O}(\log(m))$ auxiliary qubits, and may be implemented with $\mathcal{O}(m^2d^2)$ one and two-qubit gates in the worst case~\cite{nielsen2002quantum}. The Stinespring method implements the entire dynamical map in one go and therefore, one can simply proceed to measure the observable $O$. The number of shots $N$ is chosen such that $Var(\widehat{\langle O\rangle})\leq v$.

One way to possibly reduce the number of gates is by completing the Stinespring unitary to minimize the number of gates in its decomposition rather than by a naive application of Gram-Schmidt procedure. While the above analysis does not assume any knowledge about the Kraus operators, the gate decomposition complexity can also improve greatly for some sparse or specially-structured Kraus operators. 

\begin{example}\label{ex:stinespring}
    Let's consider the amplitude damping channel, with Kraus operators $A_0 = |0\rangle\langle 0| + \sqrt{1-p} |1\rangle\langle 1|$ and $A_1 = \sqrt{p} |0 \rangle\langle 1|$.  The Stinespring unitary may be filled in as follows
    \begin{minipage}{0.49\textwidth}
        \begin{align*}
            U_{\text{S}} = 
            \begin{bmatrix}
                1 & 0 & 0 & 0\\
                0 & \sqrt{1-p} & -\sqrt{p} & 0\\
                0 & \sqrt{p} & \sqrt{1-p} & 0\\
                0 & 0 & 0 & 1
            \end{bmatrix}
        \end{align*}
    \end{minipage}
    \begin{minipage}{0.49\textwidth}
        \begin{center}
            \begin{quantikz}
                \lstick{$\ket{0}$}    & \gate{R_y(\theta)}    &   \ctrl{1}  &  \trash[shorten >= 0.5em]{ } \\
                \lstick{$\ket{\psi}$} & \ctrl{-1}             & \targ{}   & \qw
            \end{quantikz}
        \end{center}
    \end{minipage}
    \newline\newline
    The channel is implemented with the aid of a single auxiliary qubit initialized in state $\ket{0}$ which is traced out at the end of the evolution. The last two columns of the unitary are easy to obtain by inspection, due to the relative sparsity of the operators, and it is also straightforward to show $\theta = 2 \sin^{-1}(\sqrt{p})$ given that $R_y(\theta) = e^{-i \theta \sigma_y/2}$.
\end{example}

\begin{example}
Simulating a Unitary Ensemble:
   This is inspired by the observation that many noise sources in controllable quantum systems stem from imperfect control, leading to a unitary implementation which can vary from realization to realization, e.g. disordered evolution \cite{kropf2016effective}. One example is that of a continuous control pulse used to enact a particular unitary gate, where the pulse itself is prone to statistical fluctuations \cite{google-flux}. Moreover, Pauli errors are captured by the same framework.  For concreteness, we consider the case {of random unitary channels \cite{Audenaert2008OnRU} also known as {\it random external fields} \cite{alicki2007quantum}}, where the quantum map is described by a discrete convex combination of unitaries $U_i$
\begin{equation}
    \Lambda(\rho) = \sum_{i=1}^m q_i U_i \rho U_i^\dag,
    \label{eq:unitary-ensemble}
\end{equation}
where $q_i>0$ are probabilities ($\sum_i q_i=1$).

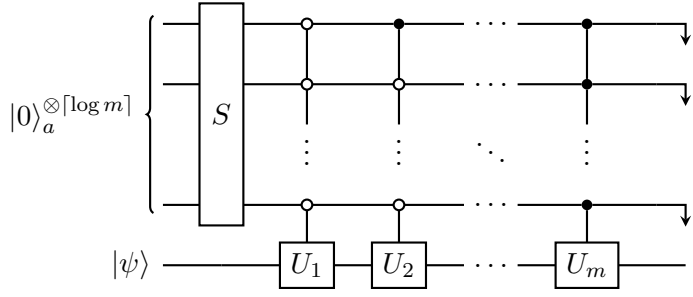
\begin{figure}[h!]
\centering
\begin{tikzpicture}
    \node[scale=1.] {
   		\begin{quantikz}[row sep={0.8cm,between origins}]
   	\lstick[4]{$\ket{0}_a^{\otimes \lceil \log m\rceil}$} & \gate[4,nwires=3]{{S}} & \octrl{1} & \ctrl{1} & \ \ldots\ \qw & \ctrl{1} & \trash[shorten >= 0.5em]{ } \\
   		    & & \octrl{1} & \octrl{1} &  \ \ldots\ \qw & \ctrl{1} & \trash[shorten >= 0.5em]{ } \\
   		    & & \ \vdots\ & \vdots & \ddots & \ \vdots \ & & \\
   		    & & \octrl{1} & \octrl{1} &  \ \ldots\ \qw & \ctrl{1} & \trash[shorten >= 0.5em]{ } \\
   		    \lstick{$\ket{\psi}$} & \qw & \gate{U_1} & \gate{U_2} & \ \ldots\ \qw & \gate{U_m} & \qw
        \end{quantikz}};
\end{tikzpicture}
\caption{Circuit construction for simulating the unitary ensemble \eqref{eq:unitary-ensemble}.  The trailing  symbols $\downarrow$ on the auxiliary wires indicate trace-out.}
\label{fig:unitary_ensemble}
\end{figure}

An illustration of the proposed scheme to enact Eq.~\eqref{eq:unitary-ensemble} is shown in Fig.~\ref{fig:unitary_ensemble}, which proceeds in two steps.  First, a state preparation unitary $S$ is implemented to prepare with $\mathcal{O}(\log m)$ auxiliary qubits the state $\sum_{i=1}^m \sqrt{q_i} |i\rangle$. Next, multi-controlled unitaries $CU_i$ are applied, conditioned on the state of the auxiliary qubits, i.e. $CU_i |i\rangle |\psi\rangle = |i\rangle U_i |\psi\rangle$.
The state of the original system plus auxiliary qubits is given by
\begin{equation}
    |\Psi\rangle := \sum_i \sqrt{q_i}\, |i\rangle \otimes  U_i |\psi\rangle \;,
    \label{eq:unitary-ensemble-full-state}
\end{equation}
where for ease of notation we assume the initial state is pure $|\psi\rangle$, though nothing in the above scheme is modified if it is an arbitrary mixed state.  Finally, by tracing out (i.e. leaving unmeasured) the auxiliary qubits, one can see Eq.~\eqref{eq:unitary-ensemble-full-state} becomes precisely Eq.~\eqref{eq:unitary-ensemble}:
\begin{equation}
    \Tr_a[|\Psi\rangle\langle \Psi|] = \Tr_a \left[\sum_{i,j} \sqrt{q_iq_j} |i\rangle \langle j | \otimes U_i |\psi\rangle \langle \psi | U_j^\dag \right] = \sum_i q_i U_i |\psi \rangle \langle \psi | U_i^\dag \;.
\end{equation}
At this point, any expectation value can be measured as desired using this output state.
\end{example}

\subsection{Parallel Simulation of a Dynamical Map}
\label{sec:BlockEncoding}
A natural attempt at overcoming the difficulty of creating and implementing the entire Stinespring unitary is to implement each Kraus operator individually in parallel~\cite{hu2020quantum}, which we will see below can reduce the gate complexity.
Since in general the Kraus operator $A_k$ is not unitary, it can itself be encoded within a larger unitary $U_k$ which can be implemented on a quantum computer. The Kraus operator can then be accessed via measurement of the auxiliary encoding qubits leading to its probabilistic implementation. We define a general block encoding as below. 
\begin{definition}[Block Encoding \cite{gilyen2019quantum}]
    Assuming that $A$ is an $n$-qubit operator\footnote{If $A$ has dimensionality $d<2^n$, then $A$ can be completed to act on $n$ qubits by padding with zeros.}, $U$ is a unitary acting on $n+\ell$ qubits, and $\alpha \in \mathbb{R}^+$ is an upper bound on the maximum singular value of $A$, then we say that $U$ is a $(\alpha, \ell, \Delta )$ block-encoding of $A$ if
   \begin{align}
       \| A - \alpha (\langle 0 |^{\otimes \ell} \otimes \mathbbm{1}) U (| 0 \rangle^{\otimes \ell} \otimes \mathbbm{1}) \| \leq \Delta \;.
   \end{align}
    Without loss of generality, we assume that $A/\alpha$ is the top-left block of $U$
   \begin{align}
    U = \begin{bmatrix}
            \quad A/\alpha  & \quad \cdot \quad \\
            \quad \cdot & \cdot
        \end{bmatrix} \;,
   \end{align}
  where $\| A/\alpha\|\leq 1$.
   \label{def:blk_enc}
\end{definition}

\begin{figure}[h!]
\centering
\begin{tikzpicture}
    \node[scale=1.] {
   		\begin{quantikz}[row sep={0.7cm,between origins}]
            \lstick[wires=2]{$\ket{0}^{\otimes \ell}$}  & \qw &\gate[wires=5,nwires=3][1cm]{U_{k}} & \qw & \meter{ } \\
             & \qw & & \qw & \meter{ } \\
             & & & & & \\
            \lstick[wires=2]{$\ket{\psi}$} & \qw & \qw & \qw & \qw \rstick[wires=2]{$\frac{A_k\ket{\psi}}{\sqrt{p_k}}$} \\
             & \qw & & \qw & \qw
        \end{quantikz}};
\end{tikzpicture}
\caption{A unitary from the set of dilation unitaries $\{  U_k \}$ such that $k \in \{1,\dots,m \}$, where each implements the state $ A_k\ket{\psi}/\sqrt{p_k}$ conditioned on measuring the auxiliary qubit in state $\ket{0}^{\otimes \ell}$.}
\label{fig:Block}
\end{figure}
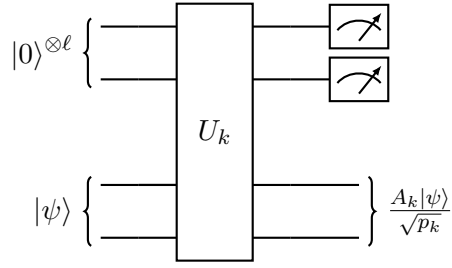

The trace preservation constraint implies that all Kraus operators are contractions such that $\| A_k\|\leq 1 \ \forall k$. Therefore they may be embedded with $\alpha=1$ for any general block encoding~\cite{hu2020quantum}.
We assume that each $A_k$ is block encoded in its respective unitary $U_k$ with $\ell$ encoding auxiliary qubits denoted by a $(1,\ell,0)$ block encoding. 
The action of each unitary on the state $\ket{\psi}$ is given as
\begin{align}
 U_k \ket{0}^{\otimes \ell}\otimes \ket{\psi} = \ket{0}^{\otimes \ell}\otimes  A_k \ket{\psi} + \ket{\Psi_\perp}\;,
\end{align}
where $(\langle{0}^{\otimes \ell}|\otimes \mathbbm{1})|\Psi_\perp\rangle =0$.
The state $ { A_k \ket{\psi}}/{\sqrt{p_{k}}}$ is successfully implemented with probability $p_{k}=\bra{\psi}A_k^\dagger A_k \ket{\psi}=\|A_k\ket{\psi}\|^2$ when the auxiliary qubits are measured in state $\ket{0}^{\otimes \ell}$. 
To implement the state $A_k\ket{\psi}/\sqrt{p_k}$ successfully with at least probability $1-\beta$, we require $\mathcal{O}(1/p_k \log(1/\beta))$ calls to the block encoding oracle $U_k$. Since we also need to prepare $\ket{\psi}$ each time after the failure, we require the same number of calls to the state preparation oracle $\mathcal{S}$.
The total number of expected calls to both $U_k$ and $\mathcal{S}$ to implement all $m$ Kraus operators in parallel is $E = \sum_{i=1}^m 1/p_k$ where $E\geq m^2$ as derived in App.~\ref{app:ExpectedRuns}. 
If we know the initial state $\ket{\psi}$, we can use amplitude amplification to achieve the square root advantage\cite{brassard1997exact,brassard2002quantum} to implement each state in $\mathcal{O}(1/\sqrt{p_k} \log(1/\beta))$ calls to $U_k$ and $\mathcal{S}$. 
Similarly the total number of expected calls to $U_k$ and $\mathcal{S}$ is $E^\prime=\sum_{k=1}^m 1/\sqrt{p_k}$  where $E^\prime \geq m^{3/2}$.

We can measure the expectation value of the observable $O$ in parallel which is given by $\langle  O \rangle = \sum_{k=1}^m \langle A_k^\dagger O A_k \rangle$. Each time any of the states $ {A_k \ket{\psi}}/{\sqrt{p_{k}}}$ is successfully implemented, we can proceed to perform a measurement of $ O$. 
The observable $O$ may be measured directly if one knows the eigenbasis of $O$, or when $O$ is provided as a linear combination of unitaries (e.g. Pauli strings) which may be measured separately and totaled (see Sect.~\ref{sec:LCU}).  More generally if no information is provided, one may instead use the Hadamard test to calculate the expectation value given a block-encoding of $O$. We detail such a calculation and its associated variance in Appendix~\ref{app:VarianceBlock}. This is repeated to obtain the expectation value of each term $\langle A_k^\dagger O A_k \rangle$ in the sum to calculate $\langle O \rangle$.
Finally a summation over the $m$ terms is performed to obtain the expectation value of $O$.  The condition $Var(\widehat{\langle  O \rangle}) \leq v$, implies that $\sum_k (p_k-\langle A_k^\dagger O A_k\rangle^2)/N_k \leq v$, where $N_k$ is the number of shots or measurements for each term $\langle A_k^\dagger O A_k \rangle$. 

A completely arbitrary block encoding $U_k$ uses $\mathcal{O}(L^2d^2)$ one and two qubit gates for its decomposition in the worst case where $L=2^\ell$. The total gate complexity is $\mathcal{O}(mL^2 d^2)$ and total number of additional qubits $\mathcal{O}(m \log(L))$ for the $m$ Krauses.
Our process of measuring each expectation value only requires the implementation of block encoding $U_k$ and the $(1,q,0)$ block encoding $U_O$ of observable $O$ using at most $\mathcal{O}((L^2+Q^2)d^2)$ one and two qubit gates for decomposition, where $Q=2^q$.
Arbitrary Kraus operators $ A_k$ and thus arbitrary unitaries require an exponential number (with respect to the number of qubits) of one and two-qubit gates to decompose. In many realistic scenarios, each $ A_k$ is a sparse matrix describing transitions between $s$ quantum states where $s = \mathcal{O}( \text{poly} \log d)$. In such cases, the number of one and two qubit gates for decomposition reduces to $\mathcal{O}(L^2s^2)$.
We discuss below that for special oracles like Sz.-Nagy and LCU encodings, this complexity can be reduced significantly.

\subsubsection{Sz.-Nagy Dilation}\label{sec:SzNagy}
The Sz.-Nagy dilation theorem guarantees a minimal block encoding of any arbitrary $d$-dimensional contraction $A$ using only one auxiliary qubit.

\begin{theorem}[Sz.-Nagy theorem~\cite{levy2014dilation, nagy2010harmonic,hu2020quantum}]
For every contraction operator $A\colon \bb(\hh)\to\bb(\hh) \ (\|A\|\le 1)$~\footnote{We use $\|A\|$ to denote the operator norm (i.e., the maximum singular value) throughout.} there exists a unitary dilation operator $U\colon \bb(\hh)\oplus \bb(\hh) \to \bb(\hh)\oplus \bb(\hh)$ where
\begin{equation}\label{eq:sz-nagy}
    U^{SN} = \begin{bmatrix}A & \sqrt{I-AA^\dag} \\\sqrt{I-A^\dag A} & -A^\dag\end{bmatrix}\;,
\end{equation}
\end{theorem}

\begin{figure}[h!]
\centering
\begin{tikzpicture}
        \node[scale=1.] {
       		\begin{quantikz}
\lstick[wires=1]{$\ket{0}$}  & \qw & \gate[3]{ U^{SN}_k} & \qw  & \meter[]{}  \\
\lstick[wires=2]{$\ket{\psi}$}  & \qw & & \qw & \qw \rstick[wires=2]{$\frac{A_k\ket{\psi}}{\sqrt{p_k}}$} \\
 & \qw & & \qw & \qw
\end{quantikz}};[column sep=0.3cm]
    \end{tikzpicture}
\caption {A unitary from the set of Sz.-Nagy's dilation unitaries $\{  U^{SN}_k \}$ such that $k \in \{1,\dots,m \}$, where each implements the state $ A_k\ket{\psi}/\sqrt{p_k}$ conditioned on measuring the auxiliary qubit in state $\ket{0}$.}
\end{figure}
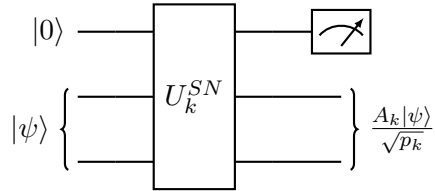

Hu, Xia and Kais~\cite{hu2020quantum} give a quantum algorithm where each Kraus operator $ A_k$ is encoded in its respective Sz.-Nagy unitary denoted by $ U^{SN}_k$ which is a (1,1,0) block encoding. The action of each unitary on a state and an auxiliary qubit is defined as
\begin{align}
 U^{SN}_k \ket{0}\otimes \ket{\psi} = \ket{0}\otimes  A_k \ket{\psi} + \ket{1}\otimes \sqrt{I-A_k^\dagger A_k}\ket{\psi}\ .
\end{align}
Since $U_k$ is a $2d$ dimensional matrix, its decomposition uses $\mathcal{O}(d^2)$ one and two qubit gates for the worst case. The total space complexity for simulating the full map of $m$ Kraus operators, in the worst case, is therefore $\mathcal{O}(md^2)$, and total number of additional qubits is $\mathcal{O}(m)$. The same results for general block encodings apply here except that the number of gates required for decomposition is $\mathcal{O}(d^2)$ which is a great reduction as it removes the dependence on $L$ which in general can depend on $d$.

Hu, Xia and Kais~\cite{hu2020quantum} provide an innovative method to calculate the expectation value 
by converting the observable $O$ to a positive-semidefinite form $\tilde{O}$ and performing Cholesky decomposition which can be incorporated into the dilation itself.  This decomposition uses at most a classical overhead of $\mathcal{O}(d^3)$ per Kraus operator. We do not require the additional step of Cholesky decomposition using our method to estimate expectation value mentioned in Sec.~\ref{sec:BlockEncoding} which reduces to $\mathcal{O}(d^2)$ complexity in the worst case given the Sz.-Nagy encodings.
Similarly as above, in the case when $A_k$ are $s$-sparse, the number of one and two qubit gates for decomposition reduces to $\mathcal{O}(s^2)$ and each Sz.-Nagy circuit becomes efficient to implement. However there are still $m$ circuits to be implemented, which can be done in parallel. 

The Sz.-Nagy dilation needs only one auxiliary qubit regardless of the dimension of the encoded $A_k$, making it the most economical block-encoding. However, constructing the dilation unitary might not be efficient as implementing the off-diagonal blocks $\sqrt{I-A^\dagger A}, \sqrt{I-AA^\dagger}$ is non-trivial. Moreover, we show in Sec~\ref{sec:Application} that the knowledge of the off-diagonal components of the Sz.-Nagy unitaries is equivalent to the knowledge of expressing any $d$-dimensional Hermitian matrix in two unitaries or a $d$-dimensional general matrix in four unitaries, instead of the $d^2$ unitaries that span the basis. 

\subsubsection{Linear Combination of Unitaries}\label{sec:LCU}

The Linear Combination of Unitaries (LCU) method allows for a probabilistic implementation of an arbitrary operator in a quantum circuit realized as a linear combination of unitaries~\cite{10.5555/2481569.2481570,low2019hamiltonian,KothariRobin_2014}.  Using the LCU method we can block encode an arbitrary matrix into a larger unitary given that we know its decomposition as a sum of unitaries.  
Let an arbitrary operator $ A_k$ be represented as a linear combination of $L$ unitaries 
\begin{align}
    A_k = \sum_{i=1}^L\alpha_{ki}  U_{ki} \;, 
\end{align}
where $\|  A_k\|\leq \alpha_k \:=\sum_{i=1}^L |\alpha_{ki}|$ and $ \alpha_{ki}\geq 0, \forall k,i$ without loss of generality. It is assumed that information about each Kraus operator $ A_k$ is provided in terms of $\alpha_{ki}$ as a list of $L$ numbers and each $ U_{ki}$ is assumed to be implemented with at most $c$ one and two qubit gates. Examples of such $U_{ki}$ include Pauli operators or more generally 1-sparse matrices.

\begin{figure}
\begin{center}
    \begin{tikzpicture}
        \node[scale=1] {
            \begin{quantikz}[row sep={0.8cm,between origins}]
                \lstick[4]{$\ket{0}_a^{\otimes \log L}$} & \gate[4,nwires=3]{B_k} & \octrl{1} & \ctrl{1} & \ \ldots\ \qw & \ctrl{1} & \gate[4,nwires=3]{B^\dagger_k} & \meter{} \\
                & & \octrl{1} & \octrl{1} &  \ \ldots\ \qw & \ctrl{1} & & \meter{}  \\
                & & \ \vdots\ & \vdots & \ddots & \ \vdots \ & & \\
                & & \octrl{1} & \octrl{1} &  \ \ldots\ \qw & \ctrl{1} & &  \meter{} \\
                \lstick{$\ket{\psi}$} & \qw & \gate{U_{k1}} & \gate{U_{k2}} & \ \ldots\ \qw & \gate{U_{kL}} & \qw \rstick{$\frac{A_k \ket{\psi}}{\sqrt{p_k}}$}
            \end{quantikz}
        };
    \end{tikzpicture}
\end{center}
\caption{Circuit construction to implement the LCU block encoding  unitaries $\{  ( B_k \otimes I)  U_k ( B_k^\dagger \otimes I) \}$ such that $k \in \{1,\dots,m \}$, where each implements the state $ A_k\ket{\psi}/\sqrt{p_k}$ conditioned on measuring the auxiliary qubit in state $\ket{0}^{\otimes \ell}$.
}
\label{fig:LCU}
\end{figure}
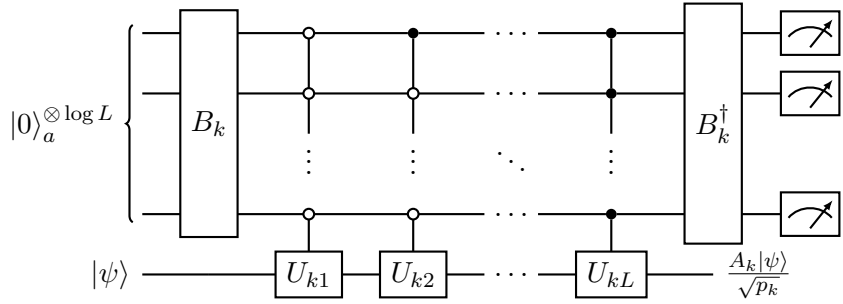

\begin{lemma}[LCU Lemma~\cite{10.5555/2481569.2481570,low2019hamiltonian}]
    Given a set of $\alpha_{ki} \geq 0$ and $U_{ki}$, with $U_{ki}U^\dagger_{ki} = I$, such that $ A_k = \sum_{i=1}^L \alpha_{ki}  U_{ki}$, there exists a quantum circuit (Fig.~\ref{fig:LCU}) that implements 
    \begin{align}\frac{ A_k}{\alpha_k}= (\bra{0}^{\otimes \ell}\otimes \mathbbm{1}) ( B_k \otimes I)  U_k ( B_k^\dagger \otimes I)(\ket{0}^{\otimes \ell} \otimes \mathbbm{1})
    \label{eq:LCU}
    \end{align}
    where $\ell = \log L$, $ B_k$ is a state preparation unitary such that $ B_k \ket{0}^{\otimes \ell} = \sum_{i=1}^L \sqrt{\frac{\alpha_{ki}}{ \alpha_k}} \ket{i}$ and $ U_k = \sum_{i=1}^L\ket{i}\bra{i}\otimes  U_{ki}$. The unitary $( B_k \otimes I)  U_k ( B_k^\dagger \otimes I)$ is a $(\alpha_k,\ell,0)$ block encoding of $A_k$.
\label{lemma:LCU}
\end{lemma} 

The LCU circuit in Fig.~\ref{fig:LCU} described by  Eq.~\eqref{eq:LCU} implements the Kraus operator $ A_k$,
requiring $\log L$ auxiliary qubits with at most $Lc$ number of decomposition gates. 
The total gate complexity is at most $mLc$ and the total number of qubits at most $m\lceil\log(L)\rceil$.
The LCU method implements the state ${{ A_k}\ket{\psi}}/({\alpha_k \sqrt{p_{k}}})$ with probability $p_{k}=\bra{\psi} A_k^\dagger  A_k \ket{\psi}/\alpha_k^2$.
The previous discussion and all results about the number of oracle calls to $U_k$ and $\mathcal{S}$ hold with the modification $A_k \rightarrow A_k/\alpha_k$, or equivalently $p_k\rightarrow p_k/\alpha_k^2$.

Cleve and Wang in Ref.~\cite{cleve_et_al:LIPIcs:2017:7477} show that all Kraus operators can be implemented in a single circuit by using an additional set of qubits applying 
the LCU method to sum over the Kraus operator index $k$.
Their method can implement $\epsilon$ approximation of a quantum channel comprising of $m$ Kraus operators in a $2^n$ dimenional space, with a quantum circuit of size $\mathcal{O}(q^2m^2  \frac{(\log(mq/\varepsilon)+n)\log(1/\varepsilon)}{(\log\log(1/\varepsilon))})$, where each Kraus is given in a linear combination of $q$ Paulis.

\section{Two-Unitary Decomposition Algorithm} \label{sec:TwoUniDecomp}
A non-unitary operator given in a block encoding unitary such as the Sz.-Nagy or LCU unitary has a probabilistic implementation. Each time the implementation fails, the block encoding unitary as well as the initial state preparation oracle must be applied again.
In this section, we provide a quantum algorithm to decompose an arbitrary $d$-dimensional contraction operator $A$ into two unitaries such that $A = (U_1+U_2)/2$ using the quantum singular value transformation (QSVT) algorithm~\cite{gilyen2019quantum,low2019hamiltonian,low2017optimal,martyn2021}. Given a state $\ket{\psi}$ prepared by $\mathcal{S}$ and a $(1,\ell,0)$ block encoding $U_A$ of an arbitrary operator $A$, the TUD algorithm implements $\tilde U_1\ket \psi,\tilde U_2\ket \psi$ such that $\|U_1-\tilde U_1\|, \|U_2-\tilde U_2\| \leq \epsilon$, with probability at least $1-\beta$ using $\mathcal{O}(\log(1/\beta))$ calls to $\mathcal{S}$, and $\mathcal{O}(1/\delta \log(1/\epsilon) \log(1/\beta))$ calls to $U_A$ and $U_A^\dagger$, where the singular values $\sigma_i $ of $A$ are such that $ \sigma_i \in [\delta,1-\delta] \ \forall i$.
Any operator can be alternatively realized by implementing the two unitaries separately with each requiring one use of the state preparation oracle $\mathcal{S}$ (i.e. a single shot of $\ket{\psi}$). This is a considerable improvement from the previously needed $\mathcal{O}(1/p \log(1/\beta))$ calls to $\mathcal{S}$. We apply the TUD algorithm to implement non-unitary Kraus operators to simulate the dynamical map and calculate the expectation value of an observable in Sec.~\ref{sec:Application}.
The queries to the block encoding unitaries $U_k$ can also be reduced significantly at the expense of an acceptable error in measurements.

{We assume access to a $(1,\ell,0)$ block encoding $U_A$ of an arbitrary operator $A = (\bra{0}^{\otimes \ell} \otimes \mathbbm{1})  U_A (\ket{0}^{\otimes \ell} \otimes \mathbbm{1})$ where $\ell = \log(L)$. Information about a sparse $A$ may typically be given in terms of its Pauli expansion, $A = \sum_{i=1}^L\alpha_{i}  P_{i} \;,$ where $P_i$ are the products of $d$-Pauli matrices. The oracle can then be created with techniques such as LCU using Ref.~\cite{10.5555/2481569.2481570,low2019hamiltonian} or Lemma~\ref{lemma:LCU} as shown in Sec.~\ref{sec:LCU}}. Mathematically, any arbitrary operator $A$ can be expressed in two unitaries as $ A=( U_1 +  U_2)/2$ which is guaranteed by the following theorem.

\begin{theorem}[Two Unitary Decomposition \cite{cui2012optimal,wu1994additive}]  \label{thm:TwoUniDecomp}
    For every contraction operator $ A\colon\bb(\mathcal{H})\rightarrow\bb(\mathcal{H}) \ (\| 
     A \| \leq 1)$, there exist two unitary operators $ U_1$ and $ U_2$ such that $ A = \frac{1}{2} ( U_1 +  U_2)$. The operators 
    $ U_1 = \sum_i   (\sigma_i+i\sqrt{1-\sigma_i^2}) \ket{w_i} \bra{v_i} $ and $ U_2 = \sum_i  (\sigma_i-i\sqrt{1-\sigma_i^2})  \ket{w_i} \bra{v_i} $
    are obtained from the singular value decomposition of $ A$: $ A = \sum_i \sigma_i \ket{w_i} \bra{v_i}$, where the $\ket{w_i}$ and $\ket{v_i}$ form orthornomal bases.
\end{theorem}

The above theorem relies on performing a SVD to obtain $ U_1$ and $ U_2$ which costs $\mathcal{O}(d^3)$ on a classical computer and is therefore {prohibitively} expensive for large quantum systems. We avoid this cost in our algorithm as we do \textit{not} perform SVD. In essence, our algorithm relies on the insight that if the SVD of $A = \sum_i \sigma_i \ket{w_i}\bra{v_i}$ then, $ U_1 =  A + i f(A)$ and $ U_2 =  A - i f(A)$, where the function $f(A) :=\sum_i \sqrt{1-\sigma_i^2} \ket{w_i}\bra{v_i}$ is defined over the singular values of $A$.

\begin{figure}
\centering
    \subfloat[]{
        \begin{tikzpicture}
\node[scale=0.74] {
\begin{quantikz}[row sep=0.22cm, column sep=0.27cm]
    \ket{0} & \gate{H} & \targ{} & \gate{e^{i\phi_n \sigma_z}} & \targ{} & \qw & \targ{} & \gate{e^{i\phi_{n-1} \sigma_z}} & \targ{} & \qw & \qw & \ \ldots\ \qw & \targ{} & \gate{e^{i\phi_1 \sigma_z}} & \targ{} & \qw & \targ{} & \gate{e^{i\phi_0 \sigma_z}} & \targ{} & \gate{H} & \qw \\
    \ket{0} & \qwbundle{\ell} & \octrl{-1} & \qw & \octrl{-1} & \gate[2]{U_A} & \octrl{-1} & \qw & \octrl{-1} & \gate[2]{U_A^\dagger} & \qw & \ \ldots\ \qw & \octrl{-1} & \qw & \octrl{-1} & \gate[2]{U^{(\dagger)}_A} & \octrl{-1} & \qw & \octrl{-1} & \qw & \qw \\
    \ket{\psi} & \qw & \qw & \qw & \qw & \qw & \qw & \qw & \qw & \qw & \qw & \ \ldots\ \qw & \qw & \qw & \qw & \qw & \qw & \qw & \qw & \qw & \qw &
\end{quantikz}
};
\end{tikzpicture}
   } \\
   \subfloat[]{
        \begin{tikzpicture}
            \node[scale=1.] {\centering
                \begin{quantikz}
                    \ket{0} & \gate[]{H} & \octrl{1} & \ctrl{1} & \gate[]{H} & \qw\\
                    \ket{0}^{\otimes \ell+1} & \qw & \gate[2,]{U_A} & \gate[2,]{ i U_{\tilde f(A)}} & \qw & \qw \\
                    \ket{\psi} & \qw & & & \qw & \qw
                \end{quantikz}
            };
        \end{tikzpicture}
    }
    \caption{(a) The QSVT circuit to implement $U_{\tilde f(A)}$ such that $\tilde f(A) = (\bra{0}^{\otimes \ell+1} \otimes \mathbbm{1})  U_{\tilde f(A)} (\ket{0}^{\otimes \ell+1} \otimes \mathbbm{1})$. The final application of the block-encoding oracle is $U_A^{(\dag)}$, which is $U_A $ or $U_A^\dagger$ for an odd or even polynomial of degree $n$ respectively. (b) Circuit $C$ in Eq.~\eqref{eq:circuit_C} to obtain the two decomposition unitaries $\tilde U_1$ and $\tilde U_2$.
    \label{fig:QSVT}}
\end{figure}
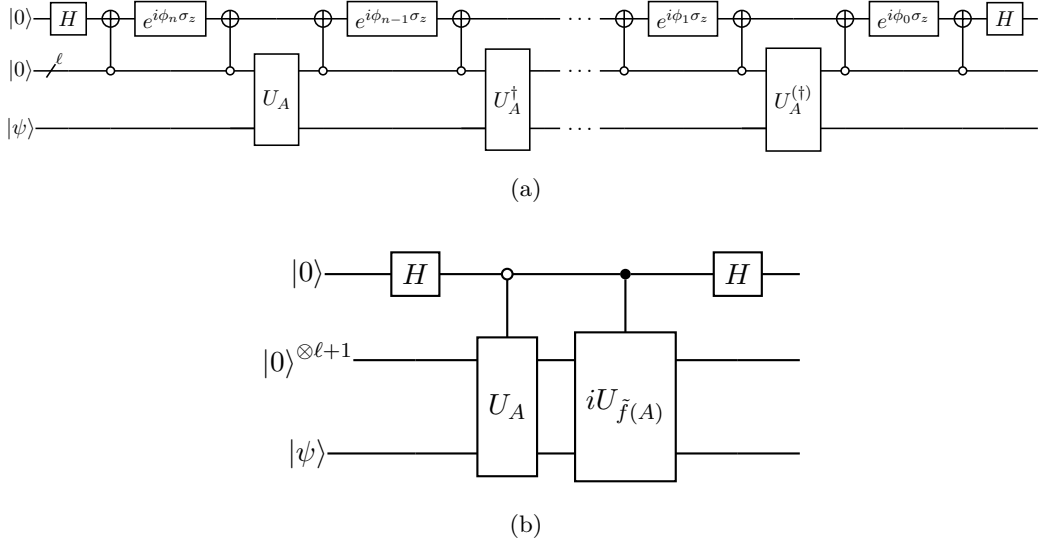

Our algorithm implements $\tilde U_1 \ket{\psi}$, $\tilde U_2 \ket{\psi}$ where $\tilde U_1 = A+i\tilde f(A)$ and $\tilde U_2 = A-i\tilde f(A)$ such that $\|U_1-\tilde U_1\|, \|U_2-\tilde U_2\| \leq \epsilon$. We create the function $\tilde f(A)$ which is an $n$-degree polynomial approximation of the function $f(A)$  such that $\|\tilde f(A) - f(A)\| \leq \epsilon$ using the QSVT algorithm as shown in Fig.~\ref{fig:QSVT} (a). The function $\tilde f(A)$ is implemented with $n/2+1$ calls to $U_A$ and $n/2$ calls to $U_A^\dagger$ such that $n$ is $\mathcal{O}({1}/{\delta}\log({1}/{\epsilon}))$ { where we assume that the singular values of $A$, $\sigma_i \in [\delta,1-\delta] \ \forall i$, such that $\delta>0$ is known.}
We then use the LCU-addition circuit in Fig.~\ref{fig:QSVT} (b) to add $A$ and $\pm i\tilde f(A)$ to obtain $\tilde U_{1,2}/2$. Using oblivious amplitude amplification (OAA) then boosts the probability of implementation arbitrarily close to one \cite{PhysRevLett.114.090502,KothariRobin_2014}, finally obtaining $\tilde U_1$ and $\tilde U_2$.

 The QSVT algorithm can implement the transformation $\tilde g(A) = \sum_i \tilde g(\sigma_i)\ket{v_i}\bra{v_i}$ for even degree polynomial of $\sigma_i$ or $\tilde f(A) = \sum_i \tilde f(\sigma_i)\ket{w_i}\bra{v_i}$ for an odd degree polynomial of $\sigma_i$. We must use an odd degree polynomial approximation as we require the basis of transformation to be  $\ket{w_i}\bra{v_i}$ to ensure that $\tilde U_{1,2}=\sum_i (\sigma_i \pm i \tilde f(\sigma_i)) \ket{w_i}\bra{v_i}$ are unitaries. Since the even function $\sqrt{1-\sigma^2}$ cannot be approximated by an odd degree polynomial, we implement the polynomial approximation of the odd function $\text{sign}(\sigma)\sqrt{1-\sigma^2}$. Both functions are identical in the relevant domain $\sigma \in (0,1]$ as singular values are always non-negative. {We present a slightly modified algorithm for the case when $\delta =0$ in the next section by implementing $A$ as a four-unitary decomposition.}

\begin{figure}[h!]
    \centering
    \includegraphics[width=\textwidth]{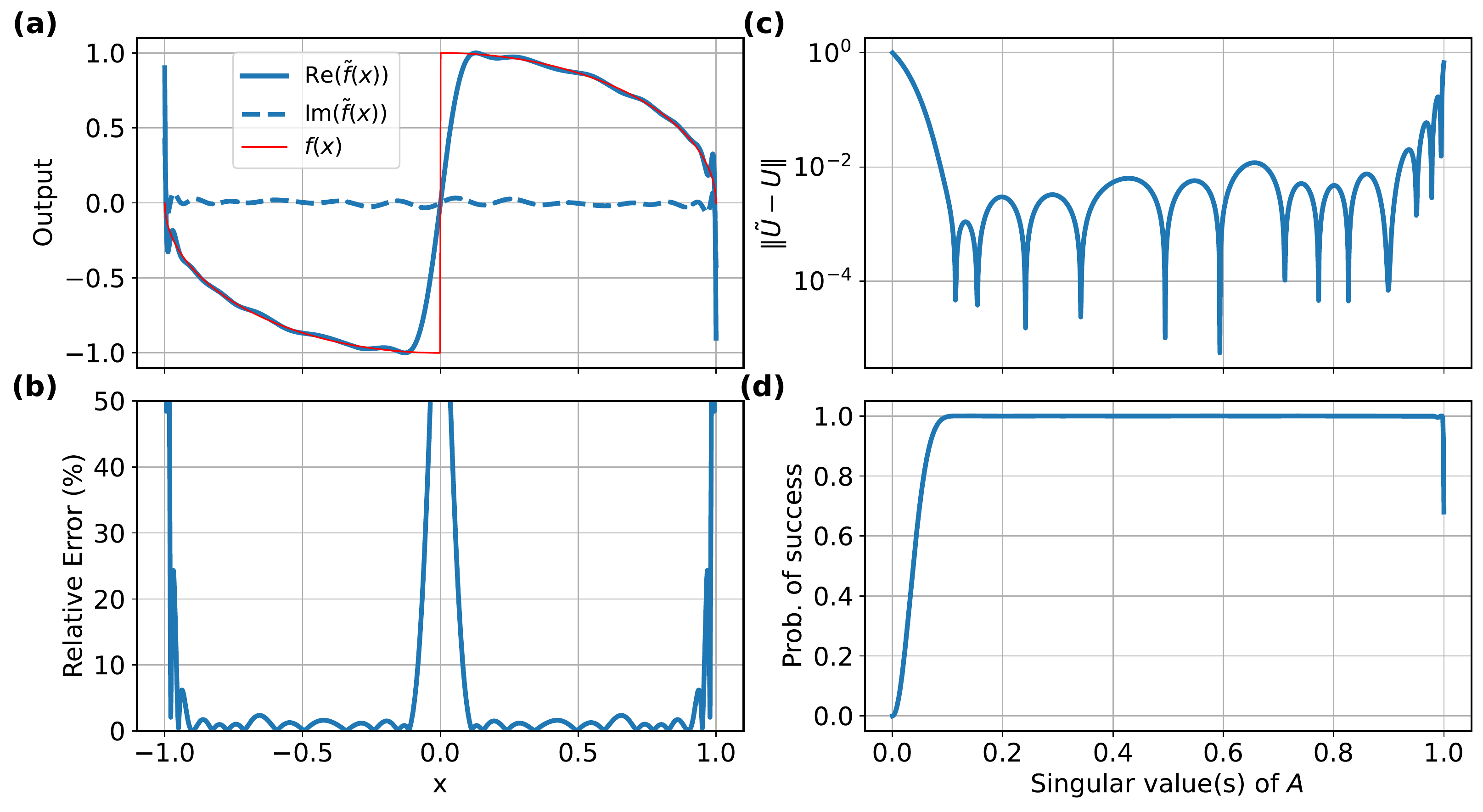}
    \caption{(\textbf{a}) Approximation of $f(x)=\text{sign}(x)\sqrt{1-x^2}$ using an odd degree-51 polynomial.  Real and imaginary components of the approximation are plotted in comparison with the ideal function. (\textbf{b}) Relative percent error between the real and ideal curves of plot (a) as a function of the scalar input.  (\textbf{c}) The operator norm (or max singular value) of the difference between the approximate QSVT unitary $\tilde{U}$ and the ideal unitary $U = A \pm if(A)$, plotted vs. the singular value(s) of the input matrix $A$. We generated 1000 two-dimensional
    random matrices 
    with both singular values being identical such that $\sigma \sim U(0,1)$. (\textbf{d}) The norm of the block-encoded matrix $\tilde{U}$ after oblivious amplitude amplification, where deviations from $1$ correspond to regions of higher error in (c).}
    \label{fig:odd_poly_51}
\end{figure}

To obtain the odd polynomial approximation of the function $f(x)=\text{sign}(x) \sqrt{1-x^2}$, we apply the QSVT transformation shown in Fig.~\ref{fig:QSVT} (a)
\begin{align}
U_{\tilde f(A)} = \Pi_{\phi_0} U_A \bigg[ \prod_{k=1}^{(n-1)/2}  \Pi_{\phi_{2k-1}} U_A^\dagger \Pi_{\phi_{2k}}  U_A \bigg] \Pi_{\phi_n}\;,
\label{eq:QSVT}
\end{align}
where $n$ is an odd number, $\Pi_\phi = e^{i\phi (2 \Pi - I)}$, $\Pi = (|0\rangle\langle 0|)^{\otimes \ell} \otimes I$, and the set of phases $\{ \phi_i\}$ in Eq.~\eqref{eq:QSVT} are efficiently computable \cite{haah2019product,chao2020finding,dong2021}.
We project it to the block-encoded subspace to obtain $\tilde f(A) = (\bra{0}^{\otimes \ell+1} \otimes \mathbbm{1})  U_{\tilde f(A)} (\ket{0}^{\otimes \ell+1} \otimes \mathbbm{1})$.
Since each application of the encoding oracle $U_A$ increases the degree of the polynomial by one, the required circuit depth such that $|f(x)-\tilde f(x)|\leq \epsilon$ where $x\in [-1+\delta,-\delta]\cup[\delta,1-\delta]$ can be found by setting degree $n=\mathcal{O}(1/\delta\log(1/\epsilon))$, see App.~\ref{app:ComplexityTUD}.
Figures~\ref{fig:odd_poly_51} (a) and (b) show the QSVT approximation for the scalar function $f(x)=\text{sign}(x) \sqrt{1-x^2}$ and the relative error using a set of 51 angles $\{ \phi_i\}$, respectively, computed using the \texttt{pyqsp} open-source repository \cite{pyqsp}.
After the previous step, we have
\begin{align}
    U_A |0\rangle^{\otimes \ell} \ket{\psi} &= |0\rangle^{\otimes \ell} A \ket{\psi} + |\Psi^\perp_{A}\rangle \label{eq:BlockEncoding} \;, \\
    U_{\tilde f(A)} |0\rangle^{\otimes \ell+1} \ket{\psi} &= |0\rangle^{\otimes \ell+1} \tilde f(A) \ket{\psi} + |\Psi^\perp_{\tilde f(A)}\rangle \;,
\end{align}
where $(\langle 0|^{\otimes \ell} \otimes \mathbbm{1})| \Psi^\perp_A \rangle =  (\langle 0|^{\otimes \ell+1} \otimes \mathbbm{1}) | \Psi^\perp_{\tilde f(A)} \rangle =0$. We add $U_A$ and $\pm i U_{\tilde f(A)}$ by implementing a LCU circuit to obtain $\tilde U_1$ and $\tilde U_2$ as shown in Fig.~\ref{fig:QSVT} (b) obtaining
\begin{align}
    C \ket{0}|0\rangle^{\otimes \ell+1}\ket{\psi} = \frac{1}{2} \{ &\ket{0} (|0\rangle^{\otimes \ell+1} \tilde U_1\ket{\psi} + |\Psi^\perp_+\rangle) \nonumber \\ + &\ket{1} (|0\rangle^{\otimes \ell+1} \tilde U_2\ket{\psi} + |\Psi^\perp_-\rangle)\}\;\label{eq:circuit_C},
\end{align}
where $|\Psi^\perp_\pm\rangle= |\Psi^\perp_A\rangle \pm i |\Psi^\perp_{\tilde f(A)}\rangle $. The decomposition unitaries $\tilde U_1$ and $\tilde U_2$ are each flagged by auxiliary qubits $\ket{0}|0\rangle^{\otimes \ell+1}$ and $\ket{1}|0\rangle^{\otimes \ell+1}$. Each of the unitaries $\tilde U_1$ or $\tilde U_2$ can now be implemented deterministically by only using oblivious amplitude amplification once~\cite{PhysRevLett.114.090502,KothariRobin_2014} for each case. We show in App.~\ref{app:probofsuccess} that the probability of successfully implementing $\tilde U_1,\tilde U_2$ is $\sim 1-3\epsilon^2 /4$. 
We shall neglect the dependence on $\epsilon^2$ moving forward by choosing it small enough. A high level description of the TUD algorithm is given in Alg.~\ref{alg:TwoUnitDecomp}.

\begin{algorithm}
\SetKwInOut{KwIn}{Input}
\SetKwInOut{KwOut}{Output}
\SetKwInOut{Runtime}{Runtime}
\SetKwInOut{Procedure}{Procedure}
\SetKwInOut{Space}{ }

\KwIn{Arbitrary matrix $A$, $\delta,\epsilon>0$ such that singular values of $A$ are $\sigma_i \in [\delta,1-\delta] \ \forall i$ block encoded in a unitary $U_A$ such that $A=(\bra{0}^{\otimes \ell} \otimes \mathbbm{1})  U_A (\ket{0}^{\otimes \ell} \otimes \mathbbm{1})$. 
The state $\ket{\psi}$ prepared by a call to the oracle $\mathcal {S}$.}
\KwOut{Flagged states $\ket{0}\ket{0}^{\otimes \ell+1}, \ket{1}\ket{0}^{\otimes \ell+1}$ indicating implementation of $ \tilde U_1 \ket{\psi}, \tilde  U_2 \ket{\psi}$ with {success} probability at least $1-\beta$ such that $\|U_1-\tilde U_1\|,\|U_2-\tilde U_2\|\leq \epsilon$ respectively where $A=\frac{1}{2}(U_1+U_2)$.}
\Runtime{$\mathcal{O}({1}/{\delta}\log({1}/{\epsilon})\log(1/\beta))$ queries to $U_A,U_A^\dagger$ using $2$ additional auxiliary qubits. $\mathcal{O}(\log(1/\beta))$ queries to the state preparation oracle~$\mathcal{S}$.}
\Procedure{} \ {Form the QSVT circuit of Eq.~\eqref{eq:QSVT} for $U_{\tilde f(A)}$ as shown in Fig.~\ref{fig:QSVT} (a)  by choosing $n=\mathcal{O}(1/\delta\log(1/\epsilon))$ to ensure $\tilde f(\sigma_i)$ is an odd 
$n$-degree $\epsilon$ approximation of the function $f(\sigma)=\text{sign}(\sigma_i)\sqrt{1-\sigma_i^2} \;, \forall i$.} \\
\ Implement the addition of $U_A$ and $\pm i U_{\tilde f(A)}$ by using the LCU circuit in Fig.~\ref{fig:QSVT} (b) and perform oblivious amplitude amplification once for each $\tilde U_1$, $ \tilde U_2$.
\caption{Two Unitary Decomposition}
\label{alg:TwoUnitDecomp}
\end{algorithm}

We demonstrate our algorithm by decomposing a set of two-dimensional matrices $A$ such that $A=U\text{diag}(\sigma,\sigma)V^\dagger$, where $U$ and $V^\dagger$ are both independently randomized\footnote{We use the \texttt{qutip} library (version 4.6.2 \cite{qutip}) and its method {\it rand\_unitary} with all default parameters which draws random Hermitian operators, multiplies them by ``$-i$'' and then exponentiates to form a unitary matrix.  We make all singular values the same so that only the error at each unique value of $\sigma$ contributes for each sampled matrix $A$.}.
We decompose $A$ and implement $\tilde U_1$ and $\tilde U_2$ by running our two-unitary decomposition algorithm for an approximating polynomial or query complexity of $n=51$.
Figure~\ref{fig:odd_poly_51} (c) shows the approximation error $\|U_1-\tilde U_1\|$ and Fig.~\ref{fig:odd_poly_51} (d) shows the probability of success of implementing $\tilde U_1$ with respect to singular values of $A$. We see that when the singular values of $A$ fall within the range $\sim [0.1,0.9]$, the approximation error remains below $10^{-2}$ and the unitaries are deterministically implemented.

The algorithm produces the correct output with  $n = \mathcal{O}( 1/\delta \log (1/\epsilon))$ uses of the oracle with the promise that singular values $\sigma_i \in [\delta,1-\delta] \ \forall i$. In practice we control $n$ and can increase it to obtain a better approximation. When given the minimum singular value $\delta$, we can increase the degree of fitting polynomial $\tilde f(x)$ in Fig.~\ref{fig:odd_poly_51} (a),(b) until $|f(x)-\tilde f(x)|\leq \epsilon, \forall x>\delta$.

The algorithm however fails when $\sigma_{min}=0$ which can happen in many realistic cases.
This problem can be addressed if we first shift all the singular values to fall within the region $[\delta^\prime ,1-\delta^\prime]$ such that $\delta^\prime > \delta$ for our choice of $\delta^\prime$,  perform the decomposition that only costs $\sim \mathcal{O}(1/\delta^\prime \log(1/\epsilon))$ and undo the shifting. The reduction in circuit depth will come at the cost of an increase in error tolerance. However, a controlled shift of this kind requires knowing the singular vectors in advance which is expensive which we therefore avoid~\footnote{ To shift 
each singular value $\sigma \rightarrow (\sigma+\lambda) /\alpha$ we have to perform the transformation $(A + \lambda WV^\dagger)/\alpha = \sum_i 
    \frac{\sigma_i + \lambda}{\alpha} |w_i\rangle\langle v_i|$ where $W,V$ are rows of left and right singular vectors respectively.}.
We show below that, with a slight modification in the formalism and by performing transformations on the eigenvalues, we can apply arbitrary eigenvalue shifts without explicit knowledge of the eigenbasis.

\subsection{Four-Unitary Decomposition}
\label{sec:FourUnitary}

\begin{algorithm}
\SetKwInOut{KwIn}{Input}
\SetKwInOut{KwOut}{Output}
\SetKwInOut{Runtime}{Runtime}
\SetKwInOut{Procedure}{Procedure}
\SetKwInOut{Space}{ }

\KwIn{Hermitian matrix $H$, ${\delta, \epsilon >0}$ with {eigenvalues of H are} $\lambda_i \in [-1+\delta,1-\delta] \ \forall i,$ block encoded in a Unitary such that $H=(\bra{0}^{\otimes \ell} \otimes \mathbbm{1})  U_H (\ket{0}^{\otimes \ell} \otimes \mathbbm{1})$. The state $\ket{\psi}$ prepared by a call to the oracle $\mathcal {S}$.}
\KwOut{Flagged states $\ket{0}\ket{0}^{\otimes \ell+1}, \ket{1}\ket{0}^{\otimes \ell+1}$ indicating implementation of $ \tilde U \ket{\psi}, \tilde  U^{\dagger} \ket{\psi}$ with probability at least $1-\beta$ such that $\|U-\tilde U\|,\|U^{\dagger}-\tilde U^{\dagger}\|\leq \epsilon$ respectively where $A=\frac{1}{2}(U+U^\dagger)$.}
\Runtime{$\mathcal{O}({1}/{\delta}\log({1}/{\epsilon})\log(1/\beta))$ queries to $U_H,U_H^\dagger$ using $2$ additional auxiliary qubits. $\mathcal{O}(\log(1/\beta))$ queries to the state preparation oracle $\mathcal{S}$.}
\Procedure{} \ {Form the QSVT circuit of Eq.~\eqref{eq:QSVT} for $U_{\tilde f(A)}$ as shown in Fig.~\ref{fig:QSVT} (a) 
by choosing $n=\mathcal{O}(1/\delta\log(1/\epsilon))$ with modification that $\tilde f(\sigma_i)$ is an even 
$n$-degree $\epsilon$ approximation of the function $f(\lambda_i)=\text{sign}(\lambda_i)\sqrt{1-\lambda_i^2} \;, \forall i$.
}. \\
\ Implement the addition of $U_H$ and $\pm i U_{\tilde f(H)}$ by using the LCU circuit in Fig.~\ref{fig:QSVT} (b) and perform oblivious amplitude amplification once for each $\tilde U$, $ \tilde U^{\dagger}$.
\caption{Two Unitary Decomposition for Hermitian Matrix}
\label{alg:TwoUnitDecompHermitian}
\end{algorithm}

\begin{figure}[t]
    \centering
    \includegraphics[width=\textwidth]{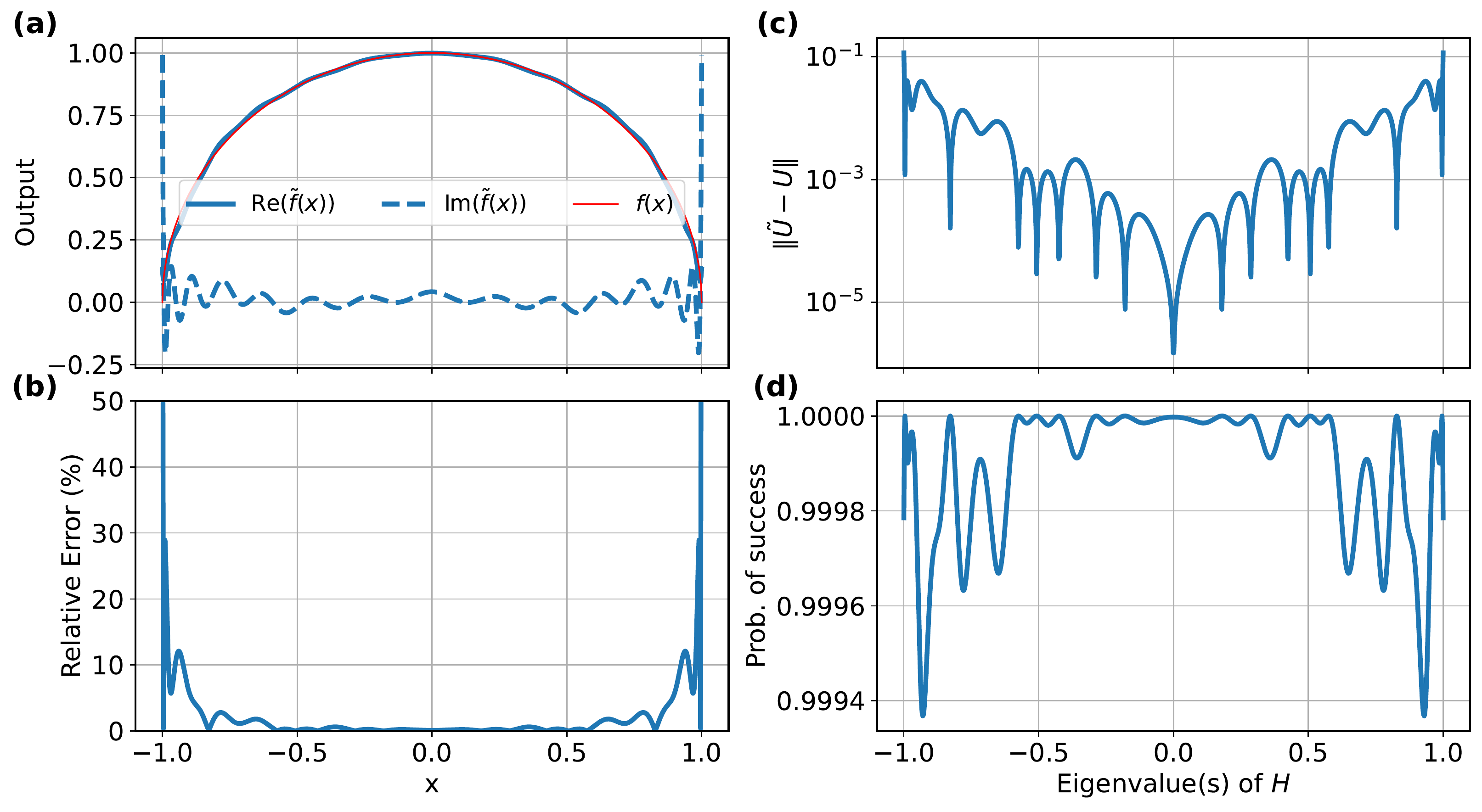}
    \caption{(\textbf{a}) Approximation of $\sqrt{1-x^2}$ using an even degree-30 polynomial.  Real and imaginary components of the approximation are plotted in comparison with the ideal function. (\textbf{b}) Relative percent error between the real and ideal curves of plot (a). as a function of the scalar input.  (\textbf{c}) The operator norm (or max singular value) of the difference between the approximate QSVT unitary $\tilde{U}$ and the ideal unitary $U = H \pm i\sqrt{1-H^2}$, plotted vs. the eigenvalue(s) of the input matrix $H$. We generated 1000 two-dimensional
    random matrices 
    with both eigenvalues being identical such that $\lambda \sim U(0,1)$. (\textbf{d}) The norm of the block-encoded matrix $\tilde{U}$ after oblivious amplitude amplification, where deviations from $1$ correspond to regions of higher error in (c).}
    \label{fig:even_poly_30}
\end{figure}

Any arbitrary matrix $A$ can be decomposed such that $A = H_1 + i H_2$ where $H_1 = (A+A^\dagger)/2$ and $H_2 = -i (A-A^\dagger)/2$ are Hermitian matrices. We show below that if we perform the two-unitary decomposition on each of the Hermitian matrices $H_1$ and $H_2$ 
such that $H_1= (U_1+U_1^\dagger)/2$ and $H_2 = (U_2 + U_2^\dagger)/2$,
we can control the shifting of the eigenvalues and therefore can control the query complexity to the oracle as desired.

A Hermitian matrix $H$ given in a $(1,\ell,0)$ block encoding $U_H$ can be decomposed into two-unitaries such that $H=(U + U^\dagger)/2$, where  $U = \sum_j (\lambda_j + i \sqrt{1-\lambda_j^2})\ket{\lambda_j}\bra{\lambda_j}$ and $\lambda_j,\ket{\lambda_j}$ are the eigenvalues and eigenvectors of $H$. We use a simplified version of our two-unitary decomposition algorithm, Alg.~\ref{alg:TwoUnitDecompHermitian}, to implement the decomposition unitaries $\tilde U,\tilde U^{\dagger}$ such that $\|U-\tilde U\|,\|U^\dagger - \tilde U^{\dagger}\| \leq \epsilon$, with at least probability $1-\beta$ using $\mathcal{O}(1/\delta \log (1/\epsilon)\log(1/\beta))$ calls to the block encoding oracle $U_H$ where the eigenvalues $\lambda_j \in [-1+\delta,1-\delta] \ \forall j$. Particularly, when given a Hermitian matrix, the QSVT algorithm will always implement both the even or odd polynomial approximations of any function in the same basis $\ket{\lambda_j}\bra{\lambda_j}$, thus removing the restriction of choosing an odd degree polynomial to ensure the unitarity of $U$. Therefore, we can choose the originally-intended even function $f(x) = \sqrt{1-x^2}$ to be approximated by an even degree polynomial $\tilde f(x)$ created by the circuit in Fig.~\ref{fig:QSVT} (a) with the final call to $U_A^\dagger$ instead of $U_A$.
Figure~\ref{fig:even_poly_30} (a),(b) show the QSVT approximation $\tilde f(x)$ for the scalar function $f(x)= \sqrt{1-x^2}$ and the relative error using a set of 30 angles $\{ \phi_i\}$. We can see that there is large error when $x\rightarrow \pm 1$.  This error can be avoided
by performing the two-unitary decomposition for $H/\alpha$; $H/\alpha = (U^{\prime}+U^{\prime\dagger})/2$ for $\alpha >1 $ to obtain $\tilde U^{\prime}$ such that $\|U^\prime - \tilde U^\prime \| \leq \epsilon$.
The effect of scaling is to compress all eigenvalues from $\lambda_i \in [-1,1] \rightarrow \lambda_i/\alpha \in [-1/\alpha,1/\alpha]$ and to avoid the error close to $-1,1$. This method thus avoids the problem when a singular value of $A$ is zero by  using a different polynomial approximation whose error does \textit{not} blow up when an eigenvalue is 0. We discuss the dependence of the query complexity on the scaling factor $\alpha$ in App.~\ref{app:FUDComplexity}.
For the example shown in Fig.~\ref{fig:even_poly_30} where $n=30$, the scaling $\alpha$ can be chosen such that the relative error shown in Fig.~\ref{fig:even_poly_30} (b) for the approximation is within the acceptable range. One could choose $\alpha= 2$ in the decomposition of $H/\alpha$, and obtain $ \|U^\prime - \tilde U^\prime\| \sim 10^{-2}$ in the scaled domain $[-1/2,1/2]$ as shown in Fig.~\ref{fig:even_poly_30} (c).

\subsection{Application: Dynamical Map Simulation}
\label{sec:Application}
{ In summary, a Kraus operator $A$ with non-zero singular values can be implemented as two separate unitaries using Alg.~\ref{alg:TwoUnitDecomp} or four unitaries if it has a vanishing singular value as shown in Sec~\ref{sec:FourUnitary} using Alg.~\ref{alg:TwoUnitDecompHermitian}. If $A$ is Hermitian, Alg.~\ref{alg:TwoUnitDecompHermitian} can be used directly to implement it as two-unitaries.}
In this subsection we show the application of the TUD algorithm to implement non-unitary Kraus operators to simulate the dynamical map and calculate expectation value of an observable. We assume access to Kraus operators in a general block encoding such as the LCU in Sec.~\ref{sec:MapGeneral}, and show the calculation for generalized amplitude damping channel using four-unitary decomposition. If one assumes access to block-encodings specifically of the Sz.-Nagy form, we show in Sec.~\ref{sec:assuming_sznagy} that the four-unitary decomposition can be obtained without using QSVT.

\subsubsection{Given General Encoding}
\label{sec:MapGeneral}
We assume access to the block encoding of $m$ Kraus operators $A_k$ such that $A_k = (\bra{0}^{\otimes \ell} \otimes \mathbbm{1})  U_{A_k} (\ket{0}^{\otimes \ell} \otimes \mathbbm{1})$ where $\ell = \log(L)$, which may be created by techniques such as LCU as shown in Sec.~\ref{sec:BlockEncoding}. 

\begin{figure}[t]
    \centering
    \includegraphics[width=\textwidth]{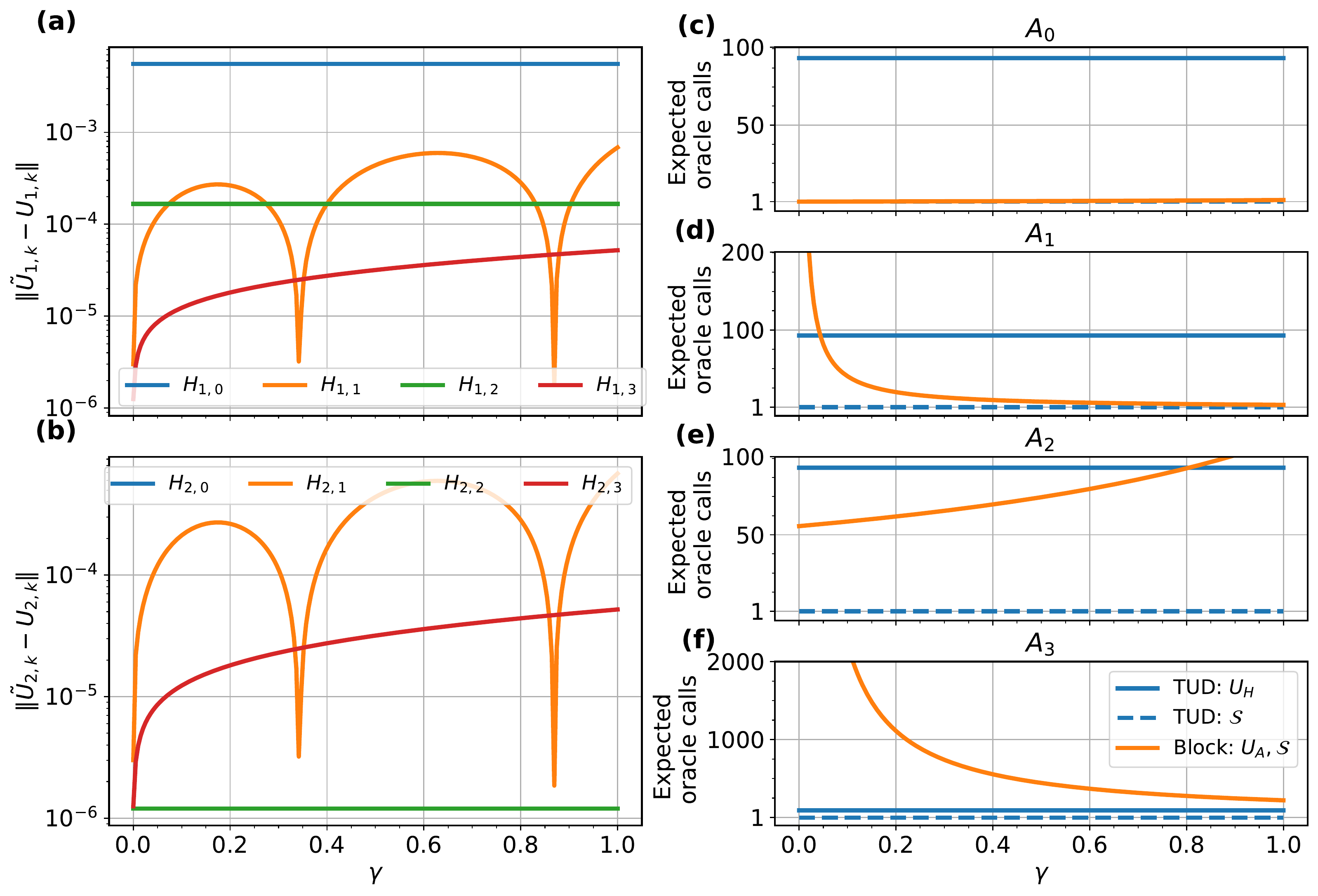}
    \caption{Error behavior and query complexity for the four Kraus operators of the generalized amplitude damping channel, as a function of the decay/excitation event probability $\gamma$.  \textbf{(a-b)} The operator norm (or max singular value) of the difference between the approximate QSVT unitary $\tilde{U}_{jk}$ and the ideal unitary $U_{jk} = H_{jk} \pm i\sqrt{1-H_{jk}^2}$, where (a) $H_{1k} = \frac{1}{2}(A_k + A_k^\dagger)$ , (b) $H_{2k} = \frac{i}{2}(A_k - A_k^\dagger)$ , and $k$ indexes the Kraus operators of the channel. \textbf{(c-f)} The expected number of calls to both the state preparation and block-encoding oracles needed to obtain a successful run for each Kraus oeprator. To implement each decomposition unitary, the number of queries to $U_H$ (either $U_{H_1},U_{H_2}$) and $\mathcal{S}$ is shown with blue solid and dashed line  respectively using the TUD method. The orange curve shows expected number of queries to both $U_A$ and $\mathcal S$ using any block encoding which is a probabilistic method.}
    \label{fig:GAD}
\end{figure}

We demonstrate our algorithm by taking the generalized amplitude damping  channel as an example. The channel describes the effect of dissipation on a single qubit at a finite temperature. Let the temperature be $T$, with the steady-state probabilities $p = e^{-E_0/kT}/\mathcal{Z}$ and $1-p =e^{-E_1/kT}/\mathcal{Z}$, where $E_0$ and $E_1$ are the energies of states $\ket{0}$ and $\ket{1}$ respectively, and where the partition function $\mathcal{Z}=e^{-E_0/kT}+e^{-E_1/kT}$. The parameter $\gamma$ controls the probability of decay which in general is a function of time. The system is described by the four Kraus operators
\begin{align}
    A_0 &=\sqrt{p} \begin{bmatrix} 1 & 0 \\ 0 & \sqrt{1-\gamma} \end{bmatrix}\;, &
    A_1 &= \sqrt{p}\begin{bmatrix} 0 & \sqrt{\gamma} \\ 0 & 0 \end{bmatrix}\;,\\
    A_2 &= \sqrt{1-p}\begin{bmatrix} \sqrt{1-\gamma} & 0 \\ 0 & 1 \end{bmatrix}\;, &
    A_3 &= \sqrt{1-p}\begin{bmatrix} 0 & 0 \\ \sqrt{\gamma} & 0 \end{bmatrix}\;.
\end{align}
We take $E_0 = 0, E_1 = 5~\text{GHz},  T=50~\text{mK}$, which are typical values for superconducting qubits resulting in $p=0.982$. Since the Kraus operators $A_1,A_3$ have a zero singular value for all values of $\gamma$, we use the four unitary decomposition to implement them. Following the discussion in Sec.~\ref{sec:FourUnitary}, each Kraus operator $A_k$ can be written as $A_k=H_{1k}+iH_{2k}$. Given $H_{1k},H_{2k}$ in the form of $(\alpha,\ell,0)$ block encodings $U_{H_{1k}},U_{H_{2k}}$, we can use the four unitary decomposition to express $A_k: A_k = \frac{\alpha}{2}(U_{1k}+U_{1k}^\dagger + i U_{2k}+ i U_{2k}^\dagger)$ where the scaling $\alpha = 1.61$ in order to restrict the input eigenvalues to a sub-region having a low error profile. We chose $n=30$ degree polynomial approximation (or equivalently the number of oracle calls) and show that the error in Fig.~\ref{fig:GAD} (a) and (b) is bounded by $10^{-2}$. Figure~\ref{fig:GAD} (c) and (d) show the expected number of oracle calls to implement the state $A_k \ket{\psi}/\sqrt{p_k}$ for $k\in\{ 0,1,2,3\}$ respectively as a function of $\gamma$. The expected number of calls to the block encoding $U_{A_{k}}$ of each Kraus operator $A_k$ as well as the the state preparation oracle $\mathcal{S}$ to prepare $\ket{\psi}$ is $\mathcal{O}(1/p_k)$ shown by the orange solid line. Instead of implementing $A_k$, if we implement the individual decomposition unitaries $\tilde U_{1k} \ket\psi, \tilde U_{1k}^\dagger \ket \psi, \tilde U_{2k} \ket\psi, \tilde U_{2k}^\dagger \ket\psi$, we see that we only require $93$ calls\footnote{We used $31$ calls to obtain $\tilde U_1/2$,  and additional $2\times31$ to perform OAA.} to the block encoding oracles $U_{H_{1k}},U_{H_{2k}}$ and one call to the state preparation oracle $\mathcal{S}$ for each which is shown by the solid and dashed blue line in Fig.~\ref{fig:GAD} (c-f) respectively.  We show that these unitaries can be used to calculate expectation values of observables.

Given that the Kraus operator $A_k$ has non-zero singular values, it can be implemented as two separate unitaries $\tilde U_{1k}, \tilde U_{2k}$ using Alg.~\ref{alg:TwoUnitDecomp}, which can be used to compute each term $\langle A_k^\dagger O A_k \rangle$ separately and then summed to obtain $\langle O\rangle$. Each term in the sum is written as
\begin{align}
   \langle \widetilde{A_k^\dagger O A_k} \rangle = \frac{1}{4}\{    \langle \tilde U_{1k}^\dagger O \tilde  U_{1k} \rangle +   \langle \tilde U_{2k}^\dagger O \tilde U_{2k} \rangle +    2\Re \langle \tilde U_{1k}^\dagger O \tilde U_{2k} \rangle\}
    \;,
    \label{eq:expect_algorithm}
\end{align}
where the first two terms are expectation values of $O$ in the states $\tilde U_{1k} \ket{\psi_i}$ and $\tilde U_{2k} \ket{\psi}$ respectively. 
The third term can be calculated by performing Hadamard test given $O$ as a string of Paulis or  block encoded in a unitary $U_O$.
We show in the App.~\ref{app:error_cancel} that the error in estimation is $|\langle\widetilde{ A_k^\dagger O A_k} \rangle - \langle{ A_k^\dagger O A_k} \rangle | \leq \epsilon$, which comes from the use of OAA and vanishes without the use of OAA. This error can be made arbitrarily small by increasing the degree of polynomial approximation by querying the block encoding $U_k$ for $\mathcal{O}(\log(1/\epsilon))$ times and therefore is ignored as it has a negligible effect in the estimation of number of runs to estimate the expectation of $\langle A^\dagger_k O A_k\rangle$.
The detailed procedure of calculation of expectation value and estimation of variance can be found in Appendix~\ref{app:VarianceTUD}.
The condition $Var\langle \hat O\rangle \leq v$, implies that 
$\sum_k(1-(\frac{1}{4}\langle U_{1k}^\dagger O U_{1k}\rangle^2 + \frac{1}{4} \langle U_{2k}^\dagger O U_{2k} \rangle^2 + \frac{1}{2} \Re\langle U_{1k}^\dagger O U_{2k}\rangle^2 ))/N_k \leq v$, where $N_k$ are the number of shots for each expectation value $\langle A_k^\dagger O A_k \rangle$. The number of runs however, is higher  than for what we proposed in probabilistic implementation of Sec.~\ref{sec:BlockEncoding}.

If the decomposition of $O$ is unknown, then it can also be further decomposed into two unitaries using the same algorithm: $O =\frac{1}{2} (U_O + U_O^{\dagger} )$. Each term can now be calculated as
\begin{align}
    \langle \widetilde{A_k^\dagger O A_k} \rangle = &\frac{1}{4} \{  \Re \langle  \tilde U_{1k}^\dagger \tilde U_O \tilde U_{1k} \rangle + \Re\langle  \tilde U_{2k}^\dagger\tilde  U_O \tilde U_{2k} \rangle 
    +\Re \langle  \tilde U_{1k}^\dagger \tilde U_O \tilde U_{2k} \rangle+\Re \langle  \tilde U_{2k}^\dagger \tilde U_{O}\tilde  U_{1k} \rangle \}
    \label{eq:expect_algorithm_O}
\end{align}
where each term in the curly bracket is calculated by a Hadamard test, thus requiring four Hadamard tests regardless of the dimension of Hilbert space.

If a singular value of a matrix is zero or very close to it, we must simulate the map by decomposing each $A_k$ into four unitaries as shown in Sec.~\ref{sec:FourUnitary}. Each expectation value is then similarly calculated
\begin{align}
 \langle\widetilde{ A_k^\dagger O A_k} \rangle &=  \frac{1}{4}\{ \langle \tilde U_{1k}^\dag O \tilde U_{1k} \rangle +\langle \tilde U_{1k} O \tilde U_{1k}^\dag \rangle  + \langle\tilde U_{2k}^\dag O\tilde U_{2k} \rangle+\langle\tilde U_{2k} O\tilde U_{2k}^\dag \rangle\}+\frac{1}{2} \Re   \langle\tilde U_{1k} O\tilde U_{1k} \rangle \nonumber \\
&+ \frac{1}{2} \Re \langle\tilde U_{2k} O\tilde U_{2k} \rangle -\frac{1}{2} \Im \{ \langle\tilde U_{1k} O\tilde U_{2k} \rangle + \langle\tilde U_{1k} O\tilde U_{2k}^\dagger \rangle + \langle\tilde U_{1k}^\dagger O\tilde U_{2k} \rangle + \langle\tilde U_{1k}^\dagger O\tilde U_{2k}^\dagger \rangle \} \;,
\label{eq:fud_expval}
\end{align}
where each term is separately computed in parallel. The first four terms in the first curly bracket can be measured directly whereas the rest can be measured via Hadamard tests.
We verify the above formula for the example of generalized amplitude damping channel by plotting the expectation value and error in App.~\ref{app:error_cancel}. 
The variance for the four-unitary case can be similarly derived to the two-unitary decomposition case presented before.

\subsubsection{Given Sz.-Nagy Encoding}\label{sec:assuming_sznagy}
\begin{figure}
\begin{center}
    \begin{tikzpicture}
        \node[scale=0.9] {
            \begin{quantikz}
                \ket{0}   & \gate[]{H} & \gate[]{e^{-i\frac{\pi}{4}\sigma_z}} & \gate[2,]{U^{SN}_{1k}} & \gate[]{e^{-i\frac{\pi}{4}\sigma_z}} & \gate[]{H} & \meter[]{}\\
                \ket{\psi}& \qwbundle{}\qw&\qw & & \qw &\qw&\qw
            \end{quantikz}
        };
    \end{tikzpicture}
\end{center}
\caption{The circuit implements $U_{1k} \ket{\psi}$ deterministically. }
\label{fig:TwoUnitSzNagy}
\end{figure}
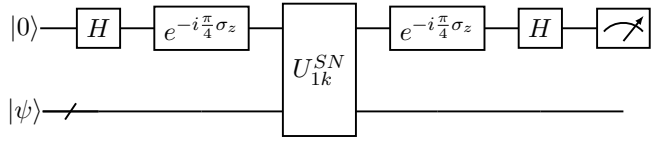
In the special case where we assume access to a minimal dilation of the Sz.-Nagy form~\cite{hu2020quantum}, we show that we do not need QSVT to obtain the two-unitary decomposition and can instead obtain it with only one call to the encoding oracle deterministically and without any error.  This assumption permits the calculation of expectation values using only a single query to the Sz.-Nagy block-encoding. 
Each $A_k = H_{1k} + i H_{2k}$ where $H_{1k} = (A_k+A_k^\dagger)/2$ and $H_{2k} = -i (A_k-A_k^\dagger)/2$ are Hermitian matrices.
We assume that both the Hermitian operators ${H}_{1k},{H}_{2k}$ are encoded in their respective Sz.-Nagy unitaries
\begin{align}
U^{SN}_{1k} = &\begin{bmatrix}
			{H}_{1k} & \sqrt{I-{H}_{1k}^2} \\
			\sqrt{I-{H}_{1k}^2} & -{H}_{1k}
			\end{bmatrix} \;,
			&
			U^{SN}_{2k} = &\begin{bmatrix}
			{H}_{2k} & \sqrt{I-{H}_{2k}^2} \\
			\sqrt{I-{H}_{2k}^2} & -{H}_{2k}
			\end{bmatrix} \;.
\end{align}
We aim to obtain the decomposition ${H}_{1k} = \frac{1}{2} (U_{1k}+U^\dagger_{1k})$. Let the eigenvalues and eigenvectors of ${H}_{1k}$ be $\lambda_{ki}, \ket{\lambda_{ki}}$, then we can see that the decomposition unitaries can be written as 
\begin{align}
    U_{1k} &= \bigoplus_i e^{i\arccos \lambda_{ki}} \ket{\lambda_{ki}}\bra{\lambda_{ki}},\\
     U^\dagger_{1k} &= \bigoplus_i e^{-i\arccos \lambda_{ki}} \ket{\lambda_{ki}}\bra{\lambda_{ki}}\;.
\end{align}
One can then easily obtain the decomposition unitaries by performing rotations on the Sz.-Nagy encoding qubit as shown in Fig.~\ref{fig:TwoUnitSzNagy}, such that 
\begin{align}
U_{1k} &= ((\bra{0} H e^{-i\frac{\pi}{4}\sigma_z}) \otimes \mathbbm{1}) U^{SN}_k ((e^{-i\frac{\pi}{4}\sigma_z} H\ket{0}) \otimes \mathbbm{1}) \;, \\
U_{1k}^{\dagger} &= ((\bra{1} H e^{-i\frac{\pi}{4}\sigma_z}) \otimes \mathbbm{1}) U^{SN}_k ((e^{-i\frac{\pi}{4}\sigma_z} H\ket{1}) \otimes \mathbbm{1})\;.
\end{align}
with $H$ being the Hadamard gate.
A similar process is done to decompose ${H}_{2k} = \frac{1}{2}(U_{2k} + U_{2k}^{\dagger})$ to obtain the decomposition of the arbitrary matrix $A_k = \frac{1}{2}(U_{1k}+U_{1k}^\dagger+ iU_{2k}+iU_{2k}^{\dagger})$. Therefore, given the Sz.-Nagy encoding, we decomposed an arbitrary operator into four unitaries without QSVT, and with only one oracle call and zero error. 
We can measure each expectation in parallel using Eq.~\eqref{eq:fud_expval} by replacing QSVT approximation unitaries $\tilde U_{ik}$ with the exact unitaries $U_{ik}$. 

Encoding the matrix in Sz.-Nagy unitary circumvents the need for QSVT due to its special structure. When given a Hermitian matrix $H$, the off-diagonal blocks of the Sz.-Nagy unitary already provide access to the transformation $f(H) = \sqrt{1-H^2}$, which can be accessed. This special structure is even present in any other general oracle but the basis required to access the off-diagonal block is unknown in general.   
Additionally, we use the four-unitary decomposition and not the two-unitary decomposition here because Sz.-Nagy dilation of $A$ has $\sqrt{I-A^\dagger A} = \sum_i \sqrt{1-\sigma^2_i} \ket{v_i} \bra{v_i}$ and $ \sqrt{I-A A^\dagger} = \sum_i \sqrt{1-\sigma^2_i} \ket{w_i} \bra{w_i}$ as the off-diagonal blocks, whereas the desired transformation is $f(A) = \sum_i \sqrt{1-\sigma^2_i} \ket{w_i} \bra{v_i}$. When $A$ is Hermitian, the Sz.-Nagy encoding provides the desired transformation as the input and output bases are the same.

\section{Conclusion}

\begin{table} 
{
\centering
\begin{tabular}{ |c|c|c|c| } 
 \hline 
 {Method} & State Preparation $\mathcal S$ & Block Encoding $U_A$ & Aux. Qubit \\ 
 \hline \hline
 Block Encoding &  &  & $\ell$  \\ 
 
LCU & $\mathcal{O}(1/{p} \log(1/\beta))$ & $\mathcal{O}(1/{p} \log(1/\beta))$ & $\ell$ \\ 
Sz.-Nagy &  &  & $1$ \\ 
 \hline
 TUD & $\mathcal{O}(\log(1/\beta))$ & $\mathcal{O}({1}/{\delta} \log (1/\epsilon) \log(1/\beta))$ & $\ell+2$\\ 
 \hline
\end{tabular}
\caption{ { Given an initial state $\ket{\psi}$ prepared by the state preparation oracle $\mathcal{S}$ and the operator $A$ block encoded in $U_A$ using $\ell$ auxiliary qubits, the first row shows the complexity of successfully implementing the state $A\ket{\psi}/\sqrt{p}$ with probability at least $1-\beta$, for general, LCU, and Sz.-Nagy block encodings. The second row  shows the complexity of implementing $\tilde U_1\ket \psi,\tilde U_2\ket \psi$ which are $\epsilon$ approximations of $U_1\ket{\psi},U_2\ket{\psi}$ such that $A=(U_1+U_2)/2$ with probability at least $1-\beta$, where the singular values of $A$ are $\sigma_i \in [\delta, 1-\delta] ,\; \forall i$.} }\label{tab:compare}}
\end{table}

The simulation of open quantum systems is challenging due to nature of the dynamical map that describes the evolution. In particular, the dynamical map can be represented in terms of Kraus operators, that, in general, fail to display special operator properties such as unitarity or hermiticity. Therefore, direct simulation on a standard (unitary based) quantum computer is impossible, i.e., to do so generally one must sacrifice some degree of accuracy, efficiency, or success probability. In order to circumvent those shortcomings, one needs to employ indirect methods that use a combination of auxiliary qubits, tracing-out or subsystem measurements. In this work, we analyzed prominent approaches such as the Stinespring dilation, the Sz.-Nagy dilation and the LCU technique. Additionally, we introduced a two-unitary decomposition algorithm as an alternative method for simulating Kraus operators, and hence, the entire dynamical map in parallel. {A comparison between the algorithms is summarized in Table}~\ref{tab:compare}. All discussed algorithms share a common feature of embedding a non-unitary Kraus operator into a larger unitary matrix. This procedure requires expanding the Hilbert space with extra auxiliary qubits to accommodate the embedding, which is usually referred to as a dilation or block encoding. 

We examined previously-introduced algorithms, that are of a probabilistic nature, by providing an estimation for both classical and quantum resources. Moreover, we introduce a quantum method to calculate the expectation value of an observable and compute the variance of its estimator, necessary for understanding the total number of runs of a given algorithm to reach certain precision. 
We provided the TUD algorithm, which compared to previous methods allowed one to express the non-unitary Kraus operator with non-zero singular values as a sum of two unitaries which can be deterministically implemented. The TUD approach relies on the quantum singular value transformation and avoids expensive $O(d^3)$ classical SVD overhead. Since each Kraus is expressed as a combination of two unitaries, one can implement each unitary independently without failure, which uses only a single access to the state preparation oracle. The two-unitary decomposition suffers a singularity in its error profile for matrices with vanishing singular values. We therefore provided a remedy in the form of a four-unitary decomposition, which decomposes a non-Hermitian Kraus operator into a sum of two Hermitian operators, that are then decomposed in two unitaries each. This procedure also allows to arbitrarily scale and shift eigenvalues. 
Even though in this paper we focused mainly on applications related to the simulation of open quantum systems, the basic algorithms discussed here rely on simulating arbitrary contractions ($\|A\| \le 1$). Thus, one can readily use the TUD algorithm in other settings that require implementing matrices without particular structure, as is often the case with classical input data accessed via an oracle. For example, methods in linear algebra such as the HHL algorithm \cite{HHL2009}, or in machine learning \cite{schuld2015introduction,biamonte2017quantum} more generally, can potentially benefit from  our approach. However, further work is required to identify in which cases the TUD algorithm can provide advantages over existing techniques.

The introduced TUD algorithm and reviewed methods are currently state-of-the-art techniques for simulation of open quantum systems. However, we treat this contribution not as the last word, but yet another step towards more efficient ways for simulating these complex systems. We hope that quantum computing can reveal new phenomena related to environment-system interactions that cannot be studied by other means - where the analytical treatment is confined to a handful of small-scale systems, and numerical techniques fail to scale with system's size. In particular, biological and chemical systems that are embedded in a non-trivial environment (a solution, a protein complex, etc.) so far have been treated with various approximation and semi-classical techniques or limited to small-size systems. Once we can tap into full scale OQS simulations, these restrictions can be lifted and a wealth of new physics (e.g. reaction mechanisms) could be potentially explored.

\section*{Acknowledgements}
NS did theory and JB did simulations, with equal contribution. We thank Sathyawageeswar Subramanian for helpful discussions on an early version of this work. 
This work was supported by the U.S. Department of Energy, Office of Science, National Quantum Information Science Research Centers, Co-design Center for Quantum Advantage (C2QA) under contract number DE- SC0012704 including through NASA-DOE interagency agreement SAA2-403601. NS, JB, SH, FW, JM are thankful for support from 
the NASA Ames Research Center, and 
NASA Academic Mission Services, Contract No. NNA16BD14C.

\bibliographystyle{unsrtnat}

\begin{thebibliography}{83}
\providecommand{\natexlab}[1]{#1}
\providecommand{\url}[1]{\texttt{#1}}
\expandafter\ifx\csname urlstyle\endcsname\relax
  \providecommand{\doi}[1]{doi: #1}\else
  \providecommand{\doi}{doi: \begingroup \urlstyle{rm}\Url}\fi

\bibitem[Manin(1980)]{manin1980computable}
Yuri Manin.
\newblock Computable and uncomputable.
\newblock \emph{Sovetskoye Radio, Moscow}, 128, 1980.

\bibitem[Feynman(1982)]{feynman1982simulating}
Richard~P. Feynman.
\newblock Simulating physics with computers.
\newblock \emph{Int. j. Theor. phys}, 21\penalty0 (6/7), 1982.
\newblock \doi{10.1007/BF02650179}.
\newblock URL \url{https://doi.org/10.1007/BF02650179}.

\bibitem[Nielsen and Chuang(2002)]{nielsen2002quantum}
Michael~A Nielsen and Isaac Chuang.
\newblock Quantum computation and quantum information, 2002.

\bibitem[Lloyd(1996)]{lloyd1996universal}
Seth Lloyd.
\newblock Universal quantum simulators.
\newblock \emph{Science}, pages 1073--1078, 1996.
\newblock \doi{10.1126/science.273.5278.1073}.
\newblock URL
  \url{https://www.science.org/doi/abs/10.1126/science.273.5278.1073}.

\bibitem[Martyn et~al.(2021)Martyn, Rossi, Tan, and Chuang]{martyn2021}
John~M. Martyn, Zane~M. Rossi, Andrew~K. Tan, and Isaac~L. Chuang.
\newblock Grand unification of quantum algorithms.
\newblock \emph{PRX Quantum}, 2:\penalty0 040203, Dec 2021.
\newblock \doi{10.1103/PRXQuantum.2.040203}.
\newblock URL \url{https://link.aps.org/doi/10.1103/PRXQuantum.2.040203}.

\bibitem[Georgescu et~al.(2014)Georgescu, Ashhab, and Nori]{RevModPhys.86.153}
I.~M. Georgescu, S.~Ashhab, and Franco Nori.
\newblock Quantum simulation.
\newblock \emph{Rev. Mod. Phys.}, 86:\penalty0 153--185, Mar 2014.
\newblock \doi{10.1103/RevModPhys.86.153}.
\newblock URL \url{https://link.aps.org/doi/10.1103/RevModPhys.86.153}.

\bibitem[Montanaro(2016)]{montanaro2016quantum}
Ashley Montanaro.
\newblock Quantum algorithms: an overview.
\newblock \emph{npj Quantum Information}, 2\penalty0 (1):\penalty0 1--8, 2016.
\newblock \doi{10.1038/npjqi.2015.23}.
\newblock URL \url{https://www.nature.com/articles/npjqi201523}.

\bibitem[Preskill(2021)]{preskill2021quantum}
John Preskill.
\newblock Quantum computing 40 years later.
\newblock \emph{arXiv:2106.10522}, 2021.
\newblock URL \url{https://arxiv.org/abs/2106.10522}.

\bibitem[Breuer et~al.(2002)Breuer, Petruccione, et~al.]{breuer2002theory}
Heinz-Peter Breuer, Francesco Petruccione, et~al.
\newblock \emph{The theory of open quantum systems}.
\newblock Oxford University Press on Demand, 2002.

\bibitem[Lindblad(1976)]{lindblad1976generators}
Goran Lindblad.
\newblock On the generators of quantum dynamical semigroups.
\newblock \emph{Communications in Mathematical Physics}, 48\penalty0
  (2):\penalty0 119--130, 1976.
\newblock \doi{10.1007/BF01608499}.

\bibitem[Gorini et~al.(1976)Gorini, Kossakowski, and
  Sudarshan]{gorini1976completely}
Vittorio Gorini, Andrzej Kossakowski, and Ennackal Chandy~George Sudarshan.
\newblock Completely positive dynamical semigroups of n-level systems.
\newblock \emph{Journal of Mathematical Physics}, 17\penalty0 (5):\penalty0
  821--825, 1976.
\newblock \doi{10.1063/1.522979}.
\newblock URL \url{https://aip.scitation.org/doi/10.1063/1.522979}.

\bibitem[Gyongyosi et~al.(2018)Gyongyosi, Imre, and
  Nguyen]{gyongyosi2018survey}
Laszlo Gyongyosi, Sandor Imre, and Hung~Viet Nguyen.
\newblock A survey on quantum channel capacities.
\newblock \emph{IEEE Communications Surveys \& Tutorials}, 20\penalty0
  (2):\penalty0 1149--1205, 2018.
\newblock \doi{10.1109/COMST.2017.2786748}.

\bibitem[Caruso et~al.(2014)Caruso, Giovannetti, Lupo, and
  Mancini]{RevModPhys.86.1203}
Filippo Caruso, Vittorio Giovannetti, Cosmo Lupo, and Stefano Mancini.
\newblock Quantum channels and memory effects.
\newblock \emph{Rev. Mod. Phys.}, 86:\penalty0 1203--1259, Dec 2014.
\newblock \doi{10.1103/RevModPhys.86.1203}.
\newblock URL \url{https://link.aps.org/doi/10.1103/RevModPhys.86.1203}.

\bibitem[Viola et~al.(1999)Viola, Knill, and Lloyd]{PhysRevLett.82.2417}
Lorenza Viola, Emanuel Knill, and Seth Lloyd.
\newblock Dynamical decoupling of open quantum systems.
\newblock \emph{Phys. Rev. Lett.}, 82:\penalty0 2417--2421, Mar 1999.
\newblock \doi{10.1103/PhysRevLett.82.2417}.
\newblock URL \url{https://link.aps.org/doi/10.1103/PhysRevLett.82.2417}.

\bibitem[Suter and \'Alvarez(2016)]{RevModPhys.88.041001}
Dieter Suter and Gonzalo~A. \'Alvarez.
\newblock Colloquium: Protecting quantum information against environmental
  noise.
\newblock \emph{Rev. Mod. Phys.}, 88:\penalty0 041001, Oct 2016.
\newblock \doi{10.1103/RevModPhys.88.041001}.
\newblock URL \url{https://link.aps.org/doi/10.1103/RevModPhys.88.041001}.

\bibitem[Magesan et~al.(2013)Magesan, Puzzuoli, Granade, and
  Cory]{PhysRevA.87.012324}
Easwar Magesan, Daniel Puzzuoli, Christopher~E. Granade, and David~G. Cory.
\newblock Modeling quantum noise for efficient testing of fault-tolerant
  circuits.
\newblock \emph{Phys. Rev. A}, 87:\penalty0 012324, Jan 2013.
\newblock \doi{10.1103/PhysRevA.87.012324}.
\newblock URL \url{https://link.aps.org/doi/10.1103/PhysRevA.87.012324}.

\bibitem[Zanardi et~al.(2016)Zanardi, Marshall, and
  Campos~Venuti]{dissi-lindblad-zanardi}
Paolo Zanardi, Jeffrey Marshall, and Lorenzo Campos~Venuti.
\newblock Dissipative universal lindbladian simulation.
\newblock \emph{Phys. Rev. A}, 93:\penalty0 022312, Feb 2016.
\newblock \doi{10.1103/PhysRevA.93.022312}.
\newblock URL \url{https://link.aps.org/doi/10.1103/PhysRevA.93.022312}.

\bibitem[\ifmmode \check{Z}\else \v{Z}\fi{}nidari\ifmmode~\check{c}\else
  \v{c}\fi{} et~al.(2010)\ifmmode \check{Z}\else
  \v{Z}\fi{}nidari\ifmmode~\check{c}\else \v{c}\fi{}, Prosen, Benenti, Casati,
  and Rossini]{PhysRevE.81.051135}
Marko \ifmmode \check{Z}\else \v{Z}\fi{}nidari\ifmmode~\check{c}\else
  \v{c}\fi{}, Toma\ifmmode \check{z}\else~\v{z}\fi{} Prosen, Giuliano Benenti,
  Giulio Casati, and Davide Rossini.
\newblock Thermalization and ergodicity in one-dimensional many-body open
  quantum systems.
\newblock \emph{Phys. Rev. E}, 81:\penalty0 051135, May 2010.
\newblock \doi{10.1103/PhysRevE.81.051135}.
\newblock URL \url{https://link.aps.org/doi/10.1103/PhysRevE.81.051135}.

\bibitem[Kastoryano and Brandao(2016)]{kastoryano2016quantum}
Michael~J Kastoryano and Fernando~GSL Brandao.
\newblock {Quantum Gibbs samplers: The commuting case}.
\newblock \emph{Communications in Mathematical Physics}, 344\penalty0
  (3):\penalty0 915--957, 2016.
\newblock \doi{10.1007/s00220-016-2641-8}.

\bibitem[Pi\ifmmode~\check{z}\else \v{z}\fi{}orn(2013)]{PhysRevA.88.043635}
Iztok Pi\ifmmode~\check{z}\else \v{z}\fi{}orn.
\newblock {One-dimensional Bose-Hubbard model far from equilibrium}.
\newblock \emph{Phys. Rev. A}, 88:\penalty0 043635, Oct 2013.
\newblock \doi{10.1103/PhysRevA.88.043635}.
\newblock URL \url{https://link.aps.org/doi/10.1103/PhysRevA.88.043635}.

\bibitem[Prosen and {\v{Z}}nidari{\v{c}}(2009)]{prosen2009matrix}
Toma{\v{z}} Prosen and Marko {\v{Z}}nidari{\v{c}}.
\newblock Matrix product simulations of non-equilibrium steady states of
  quantum spin chains.
\newblock \emph{Journal of Statistical Mechanics: Theory and Experiment},
  2009\penalty0 (02):\penalty0 P02035, 2009.
\newblock \doi{10.1088/1742-5468/2009/02/p02035}.
\newblock URL \url{https://doi.org/10.1088/1742-5468/2009/02/p02035}.

\bibitem[Prosen(2011)]{PhysRevLett.106.217206}
Toma{\v{z}} Prosen.
\newblock Open xxz spin chain: Nonequilibrium steady state and a strict bound
  on ballistic transport.
\newblock \emph{Phys. Rev. Lett.}, 106:\penalty0 217206, May 2011.
\newblock \doi{10.1103/PhysRevLett.106.217206}.
\newblock URL \url{https://link.aps.org/doi/10.1103/PhysRevLett.106.217206}.

\bibitem[Benenti et~al.(2009)Benenti, Casati, Prosen, Rossini, and
  {\v{Z}}nidari{\v{c}}]{PhysRevB.80.035110}
Giuliano Benenti, Giulio Casati, Toma{\v{z}} Prosen, Davide Rossini, and Marko
  {\v{Z}}nidari{\v{c}}.
\newblock Charge and spin transport in strongly correlated one-dimensional
  quantum systems driven far from equilibrium.
\newblock \emph{Phys. Rev. B}, 80:\penalty0 035110, Jul 2009.
\newblock \doi{10.1103/PhysRevB.80.035110}.
\newblock URL \url{https://link.aps.org/doi/10.1103/PhysRevB.80.035110}.

\bibitem[Prosen and \ifmmode \check{Z}\else
  \v{Z}\fi{}nidari\ifmmode~\check{c}\else \v{c}\fi{}(2012)]{PhysRevB.86.125118}
Toma\ifmmode \check{z}\else~\v{z}\fi{} Prosen and Marko \ifmmode \check{Z}\else
  \v{Z}\fi{}nidari\ifmmode~\check{c}\else \v{c}\fi{}.
\newblock Diffusive high-temperature transport in the one-dimensional hubbard
  model.
\newblock \emph{Phys. Rev. B}, 86:\penalty0 125118, Sep 2012.
\newblock \doi{10.1103/PhysRevB.86.125118}.
\newblock URL \url{https://link.aps.org/doi/10.1103/PhysRevB.86.125118}.

\bibitem[Huelga and Plenio(2013)]{huelga2013vibrations}
Susana~F Huelga and Martin~B Plenio.
\newblock Vibrations, quanta and biology.
\newblock \emph{Contemporary Physics}, 54\penalty0 (4):\penalty0 181--207,
  2013.
\newblock \doi{10.1080/00405000.2013.829687}.

\bibitem[Hu et~al.(2022)Hu, Head-Marsden, Mazziotti, Narang, and
  Kais]{hu2021general}
Zixuan Hu, Kade Head-Marsden, David~A. Mazziotti, Prineha Narang, and Sabre
  Kais.
\newblock A general quantum algorithm for open quantum dynamics demonstrated
  with the {F}enna-{M}atthews-{O}lson complex.
\newblock \emph{{Quantum}}, 6:\penalty0 726, May 2022.
\newblock ISSN 2521-327X.
\newblock \doi{10.22331/q-2022-05-30-726}.
\newblock URL \url{https://doi.org/10.22331/q-2022-05-30-726}.

\bibitem[Mostame et~al.(2012)Mostame, Rebentrost, Eisfeld, Kerman, Tsomokos,
  and Aspuru-Guzik]{mostame2012quantum}
Sarah Mostame, Patrick Rebentrost, Alexander Eisfeld, Andrew~J Kerman,
  Dimitris~I Tsomokos, and Al{\'a}n Aspuru-Guzik.
\newblock Quantum simulator of an open quantum system using superconducting
  qubits: exciton transport in photosynthetic complexes.
\newblock \emph{New Journal of Physics}, 14\penalty0 (10):\penalty0 105013,
  2012.
\newblock \doi{10.1088/1367-2630/14/10/105013}.
\newblock URL \url{https://doi.org/10.1088/1367-2630/14/10/105013}.

\bibitem[Sinayskiy et~al.(2012)Sinayskiy, Marais, Petruccione, and
  Ekert]{PhysRevLett.108.020602}
I.~Sinayskiy, A.~Marais, F.~Petruccione, and A.~Ekert.
\newblock Decoherence-assisted transport in a dimer system.
\newblock \emph{Phys. Rev. Lett.}, 108:\penalty0 020602, Jan 2012.
\newblock \doi{10.1103/PhysRevLett.108.020602}.
\newblock URL \url{https://link.aps.org/doi/10.1103/PhysRevLett.108.020602}.

\bibitem[Verstraete et~al.(2009)Verstraete, Wolf, and
  Cirac]{verstraete2009quantum}
Frank Verstraete, Michael~M Wolf, and J~Ignacio Cirac.
\newblock Quantum computation and quantum-state engineering driven by
  dissipation.
\newblock \emph{Nature physics}, 5\penalty0 (9):\penalty0 633--636, 2009.
\newblock \doi{10.1038/nphys1342}.
\newblock URL \url{https://www.nature.com/articles/nphys1342}.

\bibitem[Zanardi and Campos~Venuti(2014)]{coherent-ss-zanardi}
Paolo Zanardi and Lorenzo Campos~Venuti.
\newblock Coherent quantum dynamics in steady-state manifolds of strongly
  dissipative systems.
\newblock \emph{Phys. Rev. Lett.}, 113:\penalty0 240406, Dec 2014.
\newblock \doi{10.1103/PhysRevLett.113.240406}.
\newblock URL \url{https://link.aps.org/doi/10.1103/PhysRevLett.113.240406}.

\bibitem[Budich et~al.(2015)Budich, Zoller, and Diehl]{PhysRevA.91.042117}
Jan~Carl Budich, Peter Zoller, and Sebastian Diehl.
\newblock Dissipative preparation of chern insulators.
\newblock \emph{Phys. Rev. A}, 91:\penalty0 042117, Apr 2015.
\newblock \doi{10.1103/PhysRevA.91.042117}.
\newblock URL \url{https://link.aps.org/doi/10.1103/PhysRevA.91.042117}.

\bibitem[Diehl et~al.(2011)Diehl, Rico, Baranov, and Zoller]{diehl2011topology}
Sebastian Diehl, Enrique Rico, Mikhail~A Baranov, and Peter Zoller.
\newblock Topology by dissipation in atomic quantum wires.
\newblock \emph{Nature Physics}, 7\penalty0 (12):\penalty0 971--977, 2011.
\newblock \doi{10.1038/nphys2106}.
\newblock URL \url{https://www.nature.com/articles/nphys2106}.

\bibitem[Bardyn et~al.(2013)Bardyn, Baranov, Kraus, Rico, {\.I}mamo{\u{g}}lu,
  Zoller, and Diehl]{bardyn2013topology}
Charles-Edouard Bardyn, Mikhail~A Baranov, Christina~V Kraus, Enrique Rico,
  A~{\.I}mamo{\u{g}}lu, Peter Zoller, and Sebastian Diehl.
\newblock Topology by dissipation.
\newblock \emph{New Journal of Physics}, 15\penalty0 (8):\penalty0 085001,
  2013.
\newblock \doi{10.1088/1367-2630/15/8/085001}.
\newblock URL \url{https://doi.org/10.1088/1367-2630/15/8/085001}.

\bibitem[Kraus et~al.(2008)Kraus, B\"uchler, Diehl, Kantian, Micheli, and
  Zoller]{PhysRevA.78.042307}
B.~Kraus, H.~P. B\"uchler, S.~Diehl, A.~Kantian, A.~Micheli, and P.~Zoller.
\newblock Preparation of entangled states by quantum markov processes.
\newblock \emph{Phys. Rev. A}, 78:\penalty0 042307, Oct 2008.
\newblock \doi{10.1103/PhysRevA.78.042307}.
\newblock URL \url{https://link.aps.org/doi/10.1103/PhysRevA.78.042307}.

\bibitem[Reiter et~al.(2016)Reiter, Reeb, and S{\o}rensen]{reiter2016scalable}
Florentin Reiter, David Reeb, and Anders~S S{\o}rensen.
\newblock Scalable dissipative preparation of many-body entanglement.
\newblock \emph{Physical review letters}, 117\penalty0 (4):\penalty0 040501,
  2016.
\newblock \doi{10.1103/PhysRevLett.117.040501}.
\newblock URL \url{https://link.aps.org/doi/10.1103/PhysRevLett.117.040501}.

\bibitem[Kastoryano et~al.(2011)Kastoryano, Reiter, and
  S{\o}rensen]{kastoryano2011dissipative}
Michael~James Kastoryano, Florentin Reiter, and Anders~S{\o}ndberg S{\o}rensen.
\newblock Dissipative preparation of entanglement in optical cavities.
\newblock \emph{Physical review letters}, 106\penalty0 (9):\penalty0 090502,
  2011.
\newblock \doi{10.1103/PhysRevLett.106.090502}.
\newblock URL \url{https://link.aps.org/doi/10.1103/PhysRevLett.106.090502}.

\bibitem[Marshall et~al.(2019)Marshall, Campos~Venuti, and Zanardi]{q-data}
Jeffrey Marshall, Lorenzo Campos~Venuti, and Paolo Zanardi.
\newblock Classifying quantum data by dissipation.
\newblock \emph{Phys. Rev. A}, 99:\penalty0 032330, Mar 2019.
\newblock \doi{10.1103/PhysRevA.99.032330}.
\newblock URL \url{https://link.aps.org/doi/10.1103/PhysRevA.99.032330}.

\bibitem[Kliesch et~al.(2011)Kliesch, Barthel, Gogolin, Kastoryano, and
  Eisert]{kliesch2011dissipative}
Martin Kliesch, Thomas Barthel, Christian Gogolin, Michael Kastoryano, and Jens
  Eisert.
\newblock Dissipative quantum church-turing theorem.
\newblock \emph{Physical review letters}, 107\penalty0 (12):\penalty0 120501,
  2011.
\newblock \doi{10.1103/PhysRevLett.107.120501}.
\newblock URL \url{https://link.aps.org/doi/10.1103/PhysRevLett.107.120501}.

\bibitem[Wang et~al.(2011)Wang, Ashhab, and Nori]{wang2011quantum}
Hefeng Wang, Sahel Ashhab, and Franco Nori.
\newblock Quantum algorithm for simulating the dynamics of an open quantum
  system.
\newblock \emph{Physical Review A}, 83\penalty0 (6):\penalty0 062317, 2011.
\newblock \doi{10.1103/PhysRevA.83.062317}.
\newblock URL \url{https://link.aps.org/doi/10.1103/PhysRevA.83.062317}.

\bibitem[Barthel and Kliesch(2012)]{barthel2012quasilocality}
Thomas Barthel and Martin Kliesch.
\newblock Quasilocality and efficient simulation of markovian quantum dynamics.
\newblock \emph{Physical review letters}, 108\penalty0 (23):\penalty0 230504,
  2012.
\newblock \doi{10.1103/PhysRevLett.108.230504}.
\newblock URL \url{https://link.aps.org/doi/10.1103/PhysRevLett.108.230504}.

\bibitem[Han et~al.(2021)Han, Cai, Hu, Mu, Ma, Xu, Wang, Wang, Song, Zou, and
  Sun]{PhysRevLett.127.020504}
J.~Han, W.~Cai, L.~Hu, X.~Mu, Y.~Ma, Y.~Xu, W.~Wang, H.~Wang, Y.~P. Song, C.-L.
  Zou, and L.~Sun.
\newblock Experimental simulation of open quantum system dynamics via
  trotterization.
\newblock \emph{Phys. Rev. Lett.}, 127:\penalty0 020504, Jul 2021.
\newblock \doi{10.1103/PhysRevLett.127.020504}.
\newblock URL \url{https://link.aps.org/doi/10.1103/PhysRevLett.127.020504}.

\bibitem[Bacon et~al.(2001)Bacon, Childs, Chuang, Kempe, Leung, and
  Zhou]{bacon2001universal}
Dave Bacon, Andrew~M Childs, Isaac~L Chuang, Julia Kempe, Debbie~W Leung, and
  Xinlan Zhou.
\newblock Universal simulation of markovian quantum dynamics.
\newblock \emph{Physical Review A}, 64\penalty0 (6):\penalty0 062302, 2001.
\newblock \doi{10.1103/PhysRevA.64.062302}.
\newblock URL \url{https://link.aps.org/doi/10.1103/PhysRevA.64.062302}.

\bibitem[Sweke et~al.(2015)Sweke, Sinayskiy, Bernard, and
  Petruccione]{sweke2015universal}
Ryan Sweke, Ilya Sinayskiy, Denis Bernard, and Francesco Petruccione.
\newblock Universal simulation of markovian open quantum systems.
\newblock \emph{Physical Review A}, 91\penalty0 (6):\penalty0 062308, 2015.
\newblock \doi{10.1103/PhysRevA.91.062308}.
\newblock URL \url{https://link.aps.org/doi/10.1103/PhysRevA.91.062308}.

\bibitem[Hu et~al.(2020)Hu, Xia, and Kais]{hu2020quantum}
Zixuan Hu, Rongxin Xia, and Sabre Kais.
\newblock A quantum algorithm for evolving open quantum dynamics on quantum
  computing devices.
\newblock \emph{Scientific reports}, 10\penalty0 (1):\penalty0 1--9, 2020.
\newblock \doi{10.1038/s41598-020-60321-x}.
\newblock URL \url{https://www.nature.com/articles/s41598-020-60321-x}.

\bibitem[Gaikwad et~al.(2022)Gaikwad, Arvind, and Dorai]{gaikwad2022simulating}
Akshay Gaikwad, Arvind, and Kavita Dorai.
\newblock {Simulating open quantum dynamics on an NMR quantum processor using
  the Sz.-Nagy dilation algorithm}.
\newblock \emph{arXiv:2201.07687}, 2022.
\newblock URL \url{https://arxiv.org/abs/2201.07687}.

\bibitem[Head-Marsden et~al.(2021)Head-Marsden, Krastanov, Mazziotti, and
  Narang]{head2021capturing}
Kade Head-Marsden, Stefan Krastanov, David~A Mazziotti, and Prineha Narang.
\newblock {Capturing non-Markovian dynamics on near-term quantum computers}.
\newblock \emph{Physical Review Research}, 3\penalty0 (1):\penalty0 013182,
  2021.
\newblock \doi{10.1103/PhysRevResearch.3.013182}.
\newblock URL \url{https://link.aps.org/doi/10.1103/PhysRevResearch.3.013182}.

\bibitem[Childs and Wiebe(2012)]{10.5555/2481569.2481570}
Andrew~M. Childs and Nathan Wiebe.
\newblock {Hamiltonian Simulation Using Linear Combinations of Unitary
  Operations}.
\newblock \emph{Quantum Info. Comput.}, 12\penalty0 (11–12):\penalty0
  901–924, nov 2012.
\newblock ISSN 1533-7146.
\newblock \doi{10.26421/QIC12.11-12}.

\bibitem[Berry et~al.(2015)Berry, Childs, Cleve, Kothari, and
  Somma]{PhysRevLett.114.090502}
Dominic~W. Berry, Andrew~M. Childs, Richard Cleve, Robin Kothari, and
  Rolando~D. Somma.
\newblock Simulating hamiltonian dynamics with a truncated taylor series.
\newblock \emph{Phys. Rev. Lett.}, 114:\penalty0 090502, Mar 2015.
\newblock \doi{10.1103/PhysRevLett.114.090502}.
\newblock URL \url{https://link.aps.org/doi/10.1103/PhysRevLett.114.090502}.

\bibitem[Cleve and Wang(2017)]{cleve_et_al:LIPIcs:2017:7477}
Richard Cleve and Chunhao Wang.
\newblock {Efficient Quantum Algorithms for Simulating Lindblad Evolution}.
\newblock In Ioannis Chatzigiannakis, Piotr Indyk, Fabian Kuhn, and Anca
  Muscholl, editors, \emph{44th International Colloquium on Automata,
  Languages, and Programming (ICALP 2017)}, volume~80 of \emph{Leibniz
  International Proceedings in Informatics (LIPIcs)}, pages 17:1--17:14,
  Dagstuhl, Germany, 2017. Schloss Dagstuhl--Leibniz-Zentrum fuer Informatik.
\newblock ISBN 978-3-95977-041-5.
\newblock \doi{10.4230/LIPIcs.ICALP.2017.17}.
\newblock URL \url{http://drops.dagstuhl.de/opus/volltexte/2017/7477}.

\bibitem[Schlimgen et~al.(2021)Schlimgen, Head-Marsden, Sager, Narang, and
  Mazziotti]{PhysRevLett.127.270503}
Anthony~W. Schlimgen, Kade Head-Marsden, LeeAnn~M. Sager, Prineha Narang, and
  David~A. Mazziotti.
\newblock Quantum simulation of open quantum systems using a unitary
  decomposition of operators.
\newblock \emph{Phys. Rev. Lett.}, 127:\penalty0 270503, Dec 2021.
\newblock \doi{10.1103/PhysRevLett.127.270503}.
\newblock URL \url{https://link.aps.org/doi/10.1103/PhysRevLett.127.270503}.

\bibitem[Schlimgen et~al.(2022)Schlimgen, Head-Marsden, Sager-Smith, Narang,
  and Mazziotti]{schlimgen2022quantum}
Anthony~W Schlimgen, Kade Head-Marsden, LeeAnn~M Sager-Smith, Prineha Narang,
  and David~A Mazziotti.
\newblock Quantum state preparation and non-unitary evolution with diagonal
  operators.
\newblock \emph{arXiv preprint arXiv:2205.02826}, 2022.

\bibitem[Lloyd and Viola(2001)]{lloyd_viola_2001}
Seth Lloyd and Lorenza Viola.
\newblock Engineering quantum dynamics.
\newblock \emph{Phys. Rev. A}, 65:\penalty0 010101, Dec 2001.
\newblock \doi{10.1103/PhysRevA.65.010101}.
\newblock URL \url{https://link.aps.org/doi/10.1103/PhysRevA.65.010101}.

\bibitem[Shen et~al.(2017)Shen, Noh, Albert, Krastanov, Devoret, Schoelkopf,
  Girvin, and Jiang]{jiang_2017}
Chao Shen, Kyungjoo Noh, Victor~V. Albert, Stefan Krastanov, M.~H. Devoret,
  R.~J. Schoelkopf, S.~M. Girvin, and Liang Jiang.
\newblock Quantum channel construction with circuit quantum electrodynamics.
\newblock \emph{Phys. Rev. B}, 95:\penalty0 134501, Apr 2017.
\newblock \doi{10.1103/PhysRevB.95.134501}.
\newblock URL \url{https://link.aps.org/doi/10.1103/PhysRevB.95.134501}.

\bibitem[Motta et~al.(2020)Motta, Sun, Tan, O’Rourke, Ye, Minnich,
  Brand{\~a}o, and Chan]{motta2020determining}
Mario Motta, Chong Sun, Adrian~TK Tan, Matthew~J O’Rourke, Erika Ye, Austin~J
  Minnich, Fernando~GSL Brand{\~a}o, and Garnet Kin-Lic Chan.
\newblock Determining eigenstates and thermal states on a quantum computer
  using quantum imaginary time evolution.
\newblock \emph{Nature Physics}, 16\penalty0 (2):\penalty0 205--210, 2020.
\newblock \doi{10.1038/s41567-019-0704-4}.
\newblock URL \url{https://www.nature.com/articles/s41567-019-0704-4}.

\bibitem[Nishi et~al.(2021)Nishi, Kosugi, and
  Matsushita]{nishi2021implementation}
Hirofumi Nishi, Taichi Kosugi, and Yu-ichiro Matsushita.
\newblock {Implementation of quantum imaginary-time evolution method on NISQ
  devices by introducing nonlocal approximation}.
\newblock \emph{npj Quantum Information}, 7\penalty0 (1):\penalty0 1--7, 2021.
\newblock \doi{10.1038/s41534-021-00409-y}.
\newblock URL \url{https://www.nature.com/articles/s41534-021-00409-y}.

\bibitem[Sun et~al.(2021{\natexlab{a}})Sun, Motta, Tazhigulov, Tan, Chan, and
  Minnich]{PRXQuantum.2.010317}
Shi-Ning Sun, Mario Motta, Ruslan~N. Tazhigulov, Adrian~T.K. Tan, Garnet
  Kin-Lic Chan, and Austin~J. Minnich.
\newblock Quantum computation of finite-temperature static and dynamical
  properties of spin systems using quantum imaginary time evolution.
\newblock \emph{PRX Quantum}, 2:\penalty0 010317, Feb 2021{\natexlab{a}}.
\newblock \doi{10.1103/PRXQuantum.2.010317}.
\newblock URL \url{https://link.aps.org/doi/10.1103/PRXQuantum.2.010317}.

\bibitem[Sun et~al.(2021{\natexlab{b}})Sun, Shih, and Cheng]{sun2021efficient}
Shin Sun, Li-Chai Shih, and Yuan-Chung Cheng.
\newblock Efficient quantum simulation of open quantum system dynamics on noisy
  quantum computers.
\newblock \emph{arXiv preprint arXiv:2106.12882}, 2021{\natexlab{b}}.
\newblock URL \url{https://arxiv.org/abs/2106.12882}.

\bibitem[Gily{\'e}n et~al.(2019)Gily{\'e}n, Su, Low, and
  Wiebe]{gilyen2019quantum}
Andr{\'a}s Gily{\'e}n, Yuan Su, Guang~Hao Low, and Nathan Wiebe.
\newblock Quantum singular value transformation and beyond: exponential
  improvements for quantum matrix arithmetics.
\newblock In \emph{Proceedings of the 51st Annual ACM SIGACT Symposium on
  Theory of Computing}, pages 193--204, 2019.
\newblock \doi{10.1145/3313276.3316366}.
\newblock URL \url{https://dl.acm.org/doi/10.1145/3313276.3316366}.

\bibitem[Low and Chuang(2019)]{low2019hamiltonian}
Guang~Hao Low and Isaac~L Chuang.
\newblock Hamiltonian simulation by qubitization.
\newblock \emph{Quantum}, 3:\penalty0 163, 2019.
\newblock \doi{10.22331/q-2019-07-12-163}.
\newblock URL \url{https://quantum-journal.org/papers/q-2019-07-12-163/}.

\bibitem[Low and Chuang(2017{\natexlab{a}})]{low2017optimal}
Guang~Hao Low and Isaac~L Chuang.
\newblock Optimal hamiltonian simulation by quantum signal processing.
\newblock \emph{Physical Review Letters}, 118\penalty0 (1):\penalty0 010501,
  2017{\natexlab{a}}.
\newblock \doi{10.1103/PhysRevLett.118.010501}.
\newblock URL
  \url{https://journals.aps.org/prl/abstract/10.1103/PhysRevLett.118.010501}.

\bibitem[Harrow et~al.(2009)Harrow, Hassidim, and Lloyd]{HHL2009}
Aram~W. Harrow, Avinatan Hassidim, and Seth Lloyd.
\newblock Quantum algorithm for linear systems of equations.
\newblock \emph{Physical Review Letters}, 103\penalty0 (15), Oct 2009.
\newblock ISSN 1079-7114.
\newblock \doi{10.1103/physrevlett.103.150502}.
\newblock URL \url{http://dx.doi.org/10.1103/PhysRevLett.103.150502}.

\bibitem[Schuld et~al.(2015)Schuld, Sinayskiy, and
  Petruccione]{schuld2015introduction}
Maria Schuld, Ilya Sinayskiy, and Francesco Petruccione.
\newblock An introduction to quantum machine learning.
\newblock \emph{Contemporary Physics}, 56\penalty0 (2):\penalty0 172--185,
  2015.
\newblock \doi{10.1080/00107514.2014.964942}.
\newblock URL
  \url{https://www.tandfonline.com/doi/full/10.1080/00107514.2014.964942}.

\bibitem[Biamonte et~al.(2017)Biamonte, Wittek, Pancotti, Rebentrost, Wiebe,
  and Lloyd]{biamonte2017quantum}
Jacob Biamonte, Peter Wittek, Nicola Pancotti, Patrick Rebentrost, Nathan
  Wiebe, and Seth Lloyd.
\newblock Quantum machine learning.
\newblock \emph{Nature}, 549\penalty0 (7671):\penalty0 195--202, 2017.
\newblock \doi{10.1038/nature23474}.
\newblock URL \url{https://www.nature.com/articles/nature23474}.

\bibitem[Kropf et~al.(2016)Kropf, Gneiting, and
  Buchleitner]{kropf2016effective}
Chahan~M Kropf, Clemens Gneiting, and Andreas Buchleitner.
\newblock Effective dynamics of disordered quantum systems.
\newblock \emph{Physical Review X}, 6\penalty0 (3):\penalty0 031023, 2016.
\newblock \doi{10.1103/PhysRevX.6.031023}.
\newblock URL
  \url{https://journals.aps.org/prx/abstract/10.1103/PhysRevX.6.031023}.

\bibitem[McCourt et~al.(2022)McCourt, Neill, Lee, Quintana, Chen, Kelly,
  Smelyanskiy, Dykman, Korotkov, Chuang, and Petukhov]{google-flux}
Trevor McCourt, Charles Neill, Kenny Lee, Chris Quintana, Yu~Chen, Julian
  Kelly, V.~N. Smelyanskiy, M.~I. Dykman, Alexander Korotkov, Isaac~L. Chuang,
  and A.~G. Petukhov.
\newblock Learning noise via dynamical decoupling of entangled qubits.
\newblock \emph{arXiv:2201.11173}, 2022.
\newblock \doi{10.48550/ARXIV.2201.11173}.
\newblock URL \url{https://arxiv.org/abs/2201.11173}.

\bibitem[Audenaert and Scheel(2008)]{Audenaert2008OnRU}
Koenraad M.~R. Audenaert and S.~Scheel.
\newblock On random unitary channels.
\newblock \emph{New Journal of Physics}, 10:\penalty0 023011, 2008.
\newblock \doi{10.1088/1367-2630/10/2/023011}.
\newblock URL
  \url{https://iopscience.iop.org/article/10.1088/1367-2630/10/2/023011}.

\bibitem[Alicki and Lendi(2007)]{alicki2007quantum}
Robert Alicki and Karl Lendi.
\newblock \emph{Quantum dynamical semigroups and applications}, volume 717.
\newblock Springer, 2007.
\newblock \doi{10.1007/3-540-70861-8}.
\newblock URL \url{https://link.springer.com/book/10.1007/3-540-70861-8}.

\bibitem[Brassard and Hoyer(1997)]{brassard1997exact}
Gilles Brassard and Peter Hoyer.
\newblock An exact quantum polynomial-time algorithm for simon's problem.
\newblock In \emph{Proceedings of the Fifth Israeli Symposium on Theory of
  Computing and Systems}, pages 12--23. IEEE, 1997.

\bibitem[Brassard et~al.(2002)Brassard, Hoyer, Mosca, and
  Tapp]{brassard2002quantum}
Gilles Brassard, Peter Hoyer, Michele Mosca, and Alain Tapp.
\newblock Quantum amplitude amplification and estimation.
\newblock \emph{Contemporary Mathematics}, 305:\penalty0 53--74, 2002.

\bibitem[Levy and Shalit(2014)]{levy2014dilation}
Eliahu Levy and Orr~Moshe Shalit.
\newblock Dilation theory in finite dimensions: the possible, the impossible
  and the unknown.
\newblock \emph{Rocky Mountain Journal of Mathematics}, 44\penalty0
  (1):\penalty0 203--221, 2014.

\bibitem[Nagy et~al.(2010)Nagy, Foias, Bercovici, and
  K{\'e}rchy]{nagy2010harmonic}
B{\'e}la~Sz Nagy, Ciprian Foias, Hari Bercovici, and L{\'a}szl{\'o} K{\'e}rchy.
\newblock \emph{Harmonic analysis of operators on Hilbert space}.
\newblock Springer Science \& Business Media, 2010.

\bibitem[Kothari(2014)]{KothariRobin_2014}
Robin Kothari.
\newblock \emph{Efficient algorithms in quantum query complexity}.
\newblock PhD thesis, University of Waterloo, August 2014.
\newblock URL \url{http://hdl.handle.net/10012/8625}.

\bibitem[Cui et~al.(2012)Cui, Zhou, and Long]{cui2012optimal}
Jing~Xin Cui, Tao Zhou, and Gui~Lu Long.
\newblock An optimal expression of a kraus operator as a linear combination of
  unitary matrices.
\newblock \emph{Journal of Physics A: Mathematical and Theoretical},
  45\penalty0 (44):\penalty0 444011, 2012.
\newblock \doi{10.1088/1751-8113/45/44/444011}.
\newblock URL
  \url{https://iopscience.iop.org/article/10.1088/1751-8113/45/44/444011}.

\bibitem[Wu(1994)]{wu1994additive}
Pei Wu.
\newblock Additive combinations of special operators.
\newblock \emph{Banach Center Publications}, 30\penalty0 (1):\penalty0
  337--361, 1994.
\newblock URL \url{http://eudml.org/doc/262750}.

\bibitem[Haah(2019)]{haah2019product}
Jeongwan Haah.
\newblock Product decomposition of periodic functions in quantum signal
  processing.
\newblock \emph{Quantum}, 3:\penalty0 190, 2019.
\newblock \doi{10.22331/q-2019-10-07-190}.
\newblock URL \url{https://quantum-journal.org/papers/q-2019-10-07-190/}.

\bibitem[Chao et~al.(2020)Chao, Ding, Gilyen, Huang, and
  Szegedy]{chao2020finding}
Rui Chao, Dawei Ding, Andras Gilyen, Cupjin Huang, and Mario Szegedy.
\newblock Finding angles for quantum signal processing with machine precision.
\newblock \emph{arXiv preprint arXiv:2003.02831}, 2020.
\newblock URL \url{https://arxiv.org/abs/2003.02831}.

\bibitem[Dong et~al.(2021)Dong, Meng, Whaley, and Lin]{dong2021}
Yulong Dong, Xiang Meng, K.~Birgitta Whaley, and Lin Lin.
\newblock Efficient phase-factor evaluation in quantum signal processing.
\newblock \emph{Phys. Rev. A}, 103:\penalty0 042419, Apr 2021.
\newblock \doi{10.1103/PhysRevA.103.042419}.
\newblock URL \url{https://link.aps.org/doi/10.1103/PhysRevA.103.042419}.

\bibitem[Martyn et~al.()Martyn, Rossi, Tan, and Chuang]{pyqsp}
John~M. Martyn, Zane~M. Rossi, Andrew~K. Tan, and Isaac~L. Chuang.
\newblock Quantum signal processing.
\newblock \url{https://github.com/ichuang/pyqsp}.

\bibitem[Johansson et~al.(2013)Johansson, Nation, and Nori]{qutip}
J.R. Johansson, P.~D. Nation, and F.~Nori.
\newblock Qutip 2: A python framework for the dynamics of open quantum systems.
\newblock \emph{Comp. Phys. Comm.}, 184\penalty0 (1234), 2013.
\newblock \doi{10.1016/j.cpc.2012.11.019}.

\bibitem[Tong et~al.(2021)Tong, An, Wiebe, and Lin]{tong2021fast}
Yu~Tong, Dong An, Nathan Wiebe, and Lin Lin.
\newblock Fast inversion, preconditioned quantum linear system solvers, fast
  green's-function computation, and fast evaluation of matrix functions.
\newblock \emph{Physical Review A}, 104\penalty0 (3):\penalty0 032422, 2021.
\newblock \doi{10.1103/PhysRevA.104.032422}.
\newblock URL
  \url{https://journals.aps.org/pra/abstract/10.1103/PhysRevA.104.032422}.

\bibitem[Low et~al.(2016)Low, Yoder, and Chuang]{low2016}
Guang~Hao Low, Theodore~J. Yoder, and Isaac~L. Chuang.
\newblock Methodology of resonant equiangular composite quantum gates.
\newblock \emph{Physical Review X}, 6\penalty0 (4), Dec 2016.
\newblock ISSN 2160-3308.
\newblock \doi{10.1103/physrevx.6.041067}.
\newblock URL \url{http://dx.doi.org/10.1103/PhysRevX.6.041067}.

\bibitem[Low and Chuang(2017{\natexlab{b}})]{low2017hamiltonian}
Guang~Hao Low and Isaac~L Chuang.
\newblock Hamiltonian simulation by uniform spectral amplification.
\newblock \emph{arXiv preprint arXiv:1707.05391}, 2017{\natexlab{b}}.
\newblock URL \url{https://arxiv.org/abs/1707.05391}.

\bibitem[Subramanian et~al.(2019)Subramanian, Brierley, and
  Jozsa]{subramanian2019implementing}
Sathyawageeswar Subramanian, Stephen Brierley, and Richard Jozsa.
\newblock Implementing smooth functions of a hermitian matrix on a quantum
  computer.
\newblock \emph{Journal of Physics Communications}, 3\penalty0 (6):\penalty0
  065002, 2019.
\newblock \doi{10.1088/2399-6528/ab25a2}.
\newblock URL
  \url{https://iopscience.iop.org/article/10.1088/2399-6528/ab25a2}.

\end{thebibliography}

\clearpage
\appendix

\section{Total Number of Expected Runs}
\label{app:ExpectedRuns}
The number of expected runs to implement each Kraus operator $A_k$ and therefore the state $\frac{A_k\ket{\psi}}{\sqrt{p_k}} $ is $\mathcal{O}(1/p_k)$, where $p_k = \langle \psi| A_k^\dagger A_k |\psi\rangle$. We wish to find the minimum value of the total number of expected runs to implement all Kraus operators satisfying the constraint $\sum_{k=1}^m p_k = 1$. We write the cost function as 
\begin{align}
    E &= \sum_{k=1}^m \frac{1}{p_k} + \lambda \bigg(\sum_{k=1}^m p_k - 1\bigg)\;,\\
    \frac{\partial E}{ \partial p_k} &= -\frac{1}{p_k^2} + \lambda=0\;, \\
    \frac{\partial E}{ \partial\lambda} &= \sum_{k=1}^m p_k -1 = 0\;.
\end{align}
Solving the above we obtain $p_k = p_{\min} = 1/m \ \forall k$. These are the minima as $\frac{\partial^2E}{\partial p_k^2}\big|_{p_{\min}} > 0$. Putting this back in the expected number of runs, we obtain $E_{\min} = m^2$. Therefore $E\geq m^2$. When amplitude amplification is used each $A_k$ is implemented in $\mathcal{O}(1/\sqrt{p_k})$. Carrying out a similar calculation gives $E=\sum_{i=1}^m 1/\sqrt{p_k} \geq m^{3/2}$ with amplitude amplification.

\section{Error Estimation}
\tocless\subsection{Probability of Success}
\label{app:probofsuccess}
We calculate the probability of successfully implementing $\tilde U_1$ and $\tilde U_2$ after OAA.
The output of circuit $C$ in the main text Fig.~\ref{fig:QSVT} (b) is
\begin{align}
    C \ket{0}|0^{\ell+1}\rangle\ket{\psi} = \frac{1}{2} \{ &\ket{0} (|0^{\ell+1} \rangle \tilde U_1\ket{\psi} + |\Psi^\perp_+\rangle) \nonumber \\ + &\ket{1} (|0^{\ell+1} \rangle \tilde U_2\ket{\psi} + |\Psi^\perp_-\rangle)\}\;.
\end{align}
We only have to use OAA once (which calls the QSVT routine three times) to implement either $\tilde U_1$ or $\tilde U_2$, which we denote by $\tilde U$ in the calculation below. 

\begin{align}
    P_{\text{success}} &= \frac{1}{4}\|(3\tilde U^\dagger - \tilde U^\dagger \tilde U \tilde U^\dagger)(3\tilde U - \tilde U \tilde U^\dagger \tilde U)\| \\
    &= \frac{1}{4}\| \{ 9\tilde U^\dagger \tilde U - 6 (\tilde U^\dagger\tilde U)^2 +(\tilde U^\dagger \tilde U )^3 \} \| \\
    &=\frac{1}{4}\| \{ 9 (\tilde U^\dagger \tilde U-I +I) - 6 (\tilde U^\dagger \tilde U-I+I)^2 +(\tilde U^\dagger \tilde U-I+I )^3\}\|\\
    &\leq  1 - \frac{3}{4} \epsilon^2 + \mathcal{O}(\epsilon^3)
\end{align}
where we used $\| \tilde U^\dagger \tilde{U} - I\| \leq \epsilon$. This is verified in Fig.~\ref{fig:even_poly_30} (c),(d) as when $\| \tilde U - U\|\sim 10^{-2}$, the probability of success $\sim 1- 10^{-4}$.

\tocless\subsection{Simulation of Dynamical Map}\label{app:error_cancel}

The aim is to simulate the dynamical map
\begin{align} 
\Lambda(\ket{\psi}\bra{\psi}) = \sum_{k=1}^m A_k \ket{\psi} \bra{\psi} A_k^\dagger\;.
\end{align}
and calculate the expectation value
\begin{align}
\langle O \rangle = \sum_{k=1}^m \langle A_k^\dagger O A_k \rangle \;.
\end{align}
We show the calculation below for each term $\langle A_k^\dagger O A_k \rangle$ and thus remove $k$ for convenience.  We have a block-encoded operator $A$ and write its ideal decomposition as 
$A = \frac{1}{2}(U_{1}+U_{2})$. Here $U_{1}=A + i f(A)$ and $U_2 = A - if(A)$.  We can expand the expectation as
\begin{align}
    \langle A^\dagger O A \rangle = \frac{1}{4}\{ \langle U_{1}^\dagger O U_{1} \rangle +\langle U_{2}^\dagger O U_{2} \rangle + 2\Re \langle U_{1}^\dagger O U_{2} \rangle \}
    \label{eq:expect_algorithm_error}.
\end{align}

\begin{figure}
    \centering
    \includegraphics[width=0.91\textwidth]{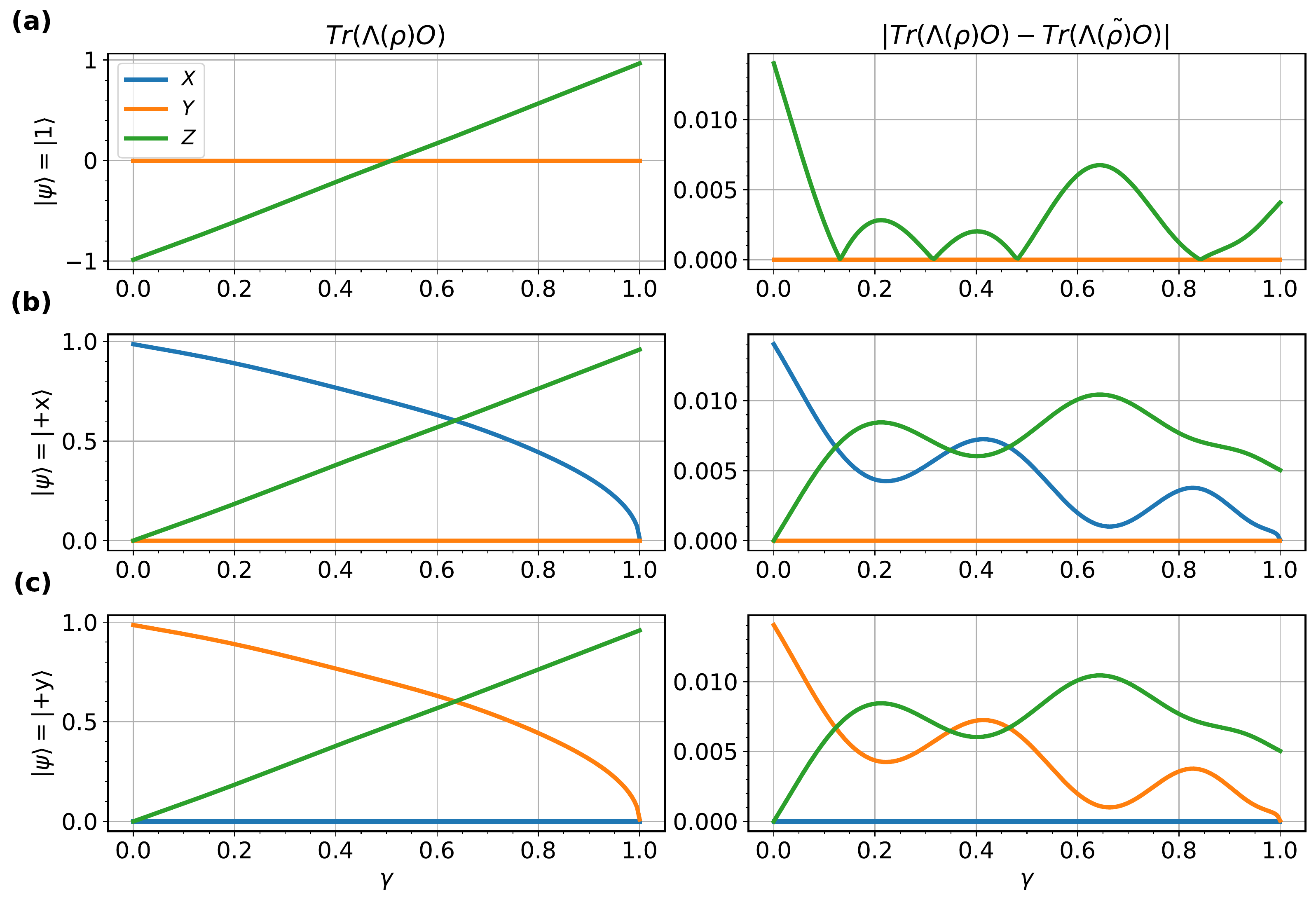}
    \caption{In row \textbf{(a)}, we plot on the left the true expectation value of a Pauli $O\in\{X,Y,Z\}$ with respect to the evolution of the state $\ket{1}$ under the generalized amplitude damping channel.  On the right, we plot the absolute difference between the true expectation value and the approximate expectation value estimated via the four-unitary decomposition of each Kraus operator $A_k(\gamma)$.  Rows \textbf{(b)} and \textbf{(c)} show the same for initial states $\ket{+x}$ and $\ket{+y}$, respectively.}
    \label{fig:GAD_error}
\end{figure}
\begin{figure}[t]
    \centering
    \includegraphics[width=\textwidth]{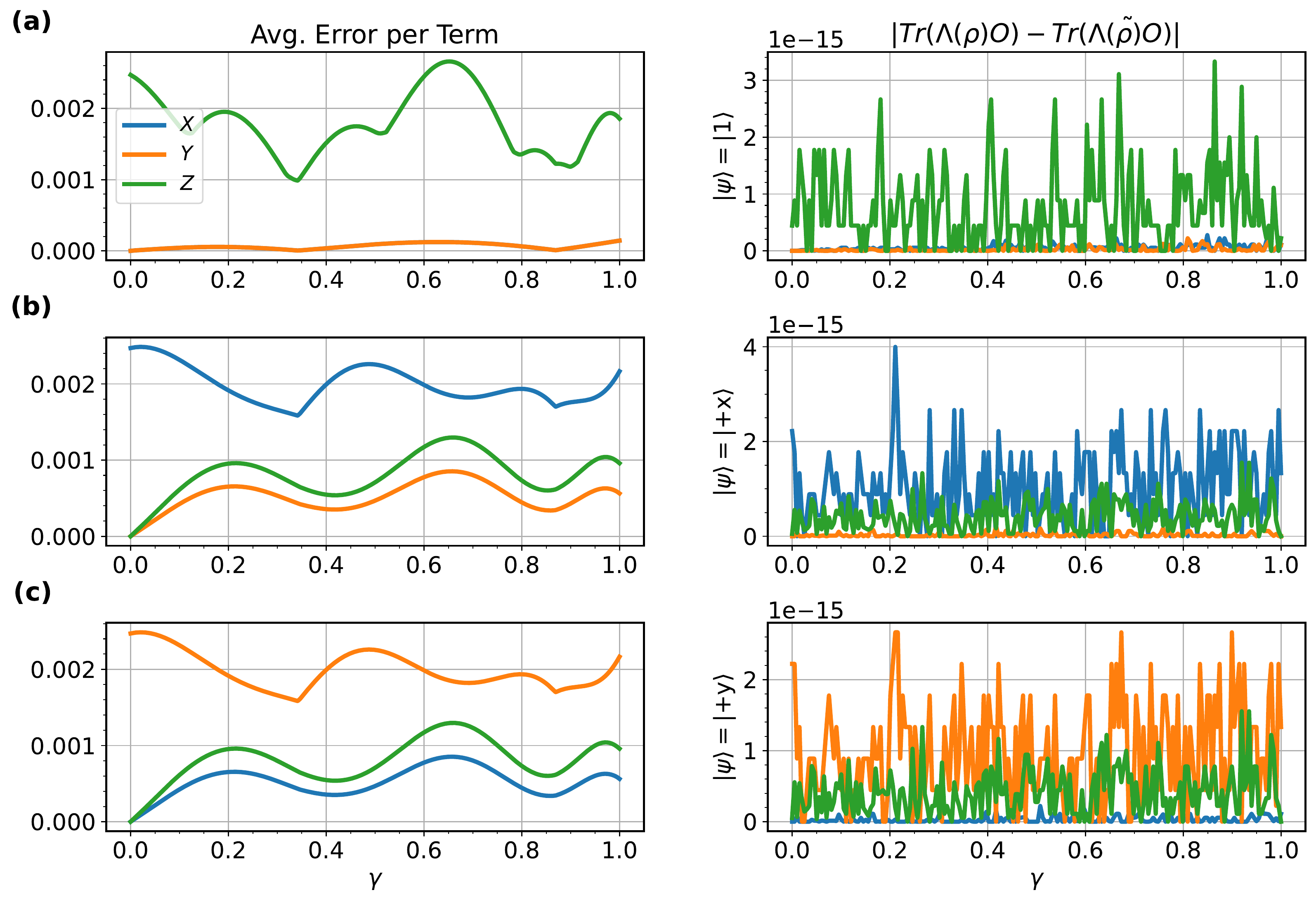}
    \caption{In row \textbf{(a)}, we plot on the right the absolute difference between the true expectation value and the approximate expectation value estimated via the four-unitary decomposition of each Kraus operator $A_k(\gamma)$ with initial state $\ket{1}$, but \textit{before OAA has been performed}.  Note the small scale of the $y$-axis.  On the left, we plot the \textit{average} error over the $10$ terms in the four-unitary decomposition of $\langle A^\dagger O A \rangle$ (see Eq.~\eqref{eq:fud_expval}).  We also average this error over the terms from the other three operators of the channel which contribute to $\Tr(\Lambda(\rho)O)$.  The scale on the right is 12 orders of magnitude larger, a clear indication that the error must cancel to obtain the scale on the left.  Rows \textbf{(b)} and \textbf{(c)} show the same for initial states $\ket{+x}$ and $\ket{+y}$, respectively.}
    \label{fig:GAD_error_per_term}
\end{figure}
In reality, we might have an error $h$ in the block encoding given by $(1,\ell,h)$ $\tilde A: \| A-\tilde A \|\leq h$. We use our two-unitary decomposition algorithm to implement the function  $\tilde f(\tilde A): \| f(\tilde A) - \tilde f (\tilde A)\| \leq \epsilon$, where $\tilde f$ is the polynomial approximation of $f$. Therefore we implement approximations of $U_{1k},U_{2k}$ which are $\tilde U_{1k}=\tilde A + i \tilde f(\tilde A) + \Delta c$ and $\tilde U_{2k}=\tilde A - i \tilde f(\tilde A) + \Delta c$, where $|\Delta c| \leq \epsilon$ comes from the error in oblivious amplitude amplification as $\tilde U_{1k},\tilde U_{2k}$ are not perfect unitaries.  We write $U_{1k}=\tilde U_{1k} + \Delta A + i \Delta f + \Delta c$ and $U_{2k}= \tilde U_{2k} + \Delta A - i \Delta f + \Delta c$ where $\Delta A = A-\tilde A$ and $\Delta f = f(A) - \tilde f (\tilde A)$. We implement ${\langle \tilde A^\dagger O \tilde A \rangle} = \frac{1}{4}\{ \langle \tilde U_{1}^\dagger O  \tilde U_{1} \rangle +\langle \tilde  U_{2}^\dagger O \tilde U_{2} \rangle + 2\Re \langle \tilde U_{1}^\dagger O \tilde  U_{2} \rangle \}$ in the circuit. Therefore we obtain 
\begin{align}
{\langle A^\dagger O A \rangle} - {\langle \widetilde{ A^\dagger O  A} \rangle} &= \frac{1}{4} \{ 2\Re \langle \tilde U_1^\dagger O (\Delta A +\Delta c + i \Delta f) \rangle + 2\Re \langle \tilde U_2^\dagger O (\Delta A + \Delta c - i \Delta f) \rangle  \nonumber \\ 
&+ 2 \Re \langle \tilde U_1^\dagger O (\Delta A + \Delta c - i\Delta f)\rangle + 2 \Re \langle \tilde U_2^\dagger O (\Delta A + \Delta c + i \Delta f)\rangle  \;,
\end{align}
where we kept the leading order terms in error.
\begin{align}
 {\langle A^\dagger O A \rangle} - {\langle \widetilde{ A^\dagger O  A} \rangle}  =  \Re \langle (\tilde U_1^\dag + \tilde U_2^\dag) O (\Delta A + \Delta c)\rangle   = 2 \Re \langle \tilde A^\dag O (\Delta A + \Delta c) \rangle
\end{align}
For each $k$, we have ${\langle A_k^\dagger O A_k \rangle} - {\langle\widetilde{  A_k^\dagger O A_k }\rangle} = 2 \Re \langle \tilde A_k^\dag O (\Delta A_k + \Delta c) \rangle$. Taking the norm,
\begin{align}
|  \langle A_k^\dagger O A_k\rangle - \langle{ \widetilde{ A_k^\dagger O  A_k} }\rangle  | = |2  \Re \langle \tilde A_k^\dag O (\Delta A_k + \Delta c)  \rangle| \leq 2  (|\langle \tilde A_k^\dag O \Delta A_k \rangle | +|\langle \tilde A_k^\dag O \Delta c \rangle | )\;,
\end{align}
and using the Cauchy-Schwarz inequality,
\begin{align}
\langle \tilde A_k^\dag O \Delta A_k \rangle &= \bra{\psi} \tilde A_k^\dag O \Delta A_k \ket{\psi} \leq \sqrt{\bra{\psi} \tilde A_k^\dagger O^2 \tilde A_k \ket{\psi} } \sqrt{\bra{\psi} \Delta A_k^\dagger \Delta A_k \ket{\psi}} \leq h \\
\langle \tilde A_k^\dag O \Delta c \rangle &= \bra{\psi} \tilde A_k^\dag O \Delta c \ket{\psi} \leq \sqrt{\bra{\psi} \tilde A_k^\dagger O^2 \tilde A_k \ket{\psi} } \sqrt{\bra{\psi} \Delta c^\dagger \Delta c \ket{\psi}} \leq \epsilon
\end{align}
 since $\bra{\psi} \tilde A_k^\dagger O^2 \tilde A_k \ket{\psi} \leq 1$ as all operators are contractions. Therefore, we obtain that 
$
| \langle A_k^\dagger O A_k\rangle - \langle{\widetilde{ A_k^\dagger O  A_k}} \rangle  | \leq 2 (\epsilon + h)
$.
We can see that when block encoding error $h=0$, the only error $\epsilon$ is due to the oblivious amplitude amplification. To be clear, the error $\Delta f$ induced by QSVT itself cancels out leaving only the error from oblivious amplitude amplification (which also originates from the QSVT error). 

A similar calculation applies when we have the four-unitary decomposition instead of the two-unitary decomposition shown above.
We calculate the expectation value and error for the example of generalized amplitude damping channel shown in the main text. Each term can be evaluated in parallel using the four unitary decomposition as follows
\begin{align}
 \langle\widetilde{ A_k^\dagger O A_k} \rangle &=  \frac{1}{4}\{ \langle \tilde U_{1k}^\dag O \tilde U_{1k} \rangle +\langle \tilde U_{1k} O \tilde U_{1k}^\dag \rangle  + \langle\tilde U_{2k}^\dag O\tilde U_{2k} \rangle+\langle\tilde U_{2k} O\tilde U_{2k}^\dag \rangle\}+\frac{1}{2} \Re   \langle\tilde U_{1k} O\tilde U_{1k} \rangle \nonumber \\
&+ \frac{1}{2} \Re \langle\tilde U_{2k} O\tilde U_{2k} \rangle -\frac{1}{2} \Im \{ \langle\tilde U_{1k} O\tilde U_{2k} \rangle + \langle\tilde U_{1k} O\tilde U_{2k}^\dagger \rangle + \langle\tilde U_{1k}^\dagger O\tilde U_{2k} \rangle + \langle\tilde U_{1k}^\dagger O\tilde U_{2k}^\dagger \rangle \} \;.
\end{align}
When OAA is used, the error $|\langle A_k^\dagger O A_k \rangle - \alpha^2 \langle \widetilde{A_k^\dagger O A_k} \rangle | \leq \alpha^2 \epsilon$, which vanishes without OAA. Figure~\ref{fig:GAD_error} (a) shows the total expectation value $\langle O \rangle$ plotted on the left calculated using the four-unitary decomposition with the initial state $\ket \psi= \ket 1$. The observable $O$ is taken as $X,Y,Z$ denoted by blue, orange and green lines respectively. The error in estimation is shown on the right for an infinite number of measurements which remains below $\approx 10^{-2}$. Similar calculation is shown for different initial states $\ket{+x},\ket{+y}$ in Fig.~\ref{fig:GAD_error} (b), (c).
We plot the error in Fig.~\ref{fig:GAD_error_per_term} and show the calculation of expectation values without performing oblivious amplitude amplification to show that it cancels for each term $\langle A_k^\dagger O A_k \rangle $.

\section{Measuring Expectation Value: Hadamard Test and Variance}
\label{app:MeasureExpect}

\begin{figure}[h!]
\begin{center}
    \begin{tikzpicture}
        \node[scale=1] {
            \begin{quantikz}
 \lstick{$\ket{0}$} & \gate{H} & \ctrl{1}  & \gate{H} & \meter{}\\
     \lstick{$\ket{\Psi}$} & \qw & \gate[1][1.cm]{U} & \qw & \qw
            \end{quantikz}
        };
    \end{tikzpicture}
\end{center}
\caption{Hadamard test}
\label{fig:htest}
\end{figure}
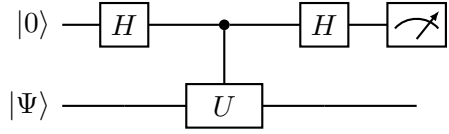
To calculate the expectation value $\Re\bra{\Psi}U\ket{\Psi}$, we perform the Hadamard test~\cite{tong2021fast} shown in Fig.~\ref{fig:htest}, denoted by the circuit $HC$.
The output of the Hadamard test circuit $HC$ is
\begin{align}
    HC \ket{0} \ket{\Psi} = \frac{1}{2}\{ \ket{0}\otimes (I+U)\ket{\Psi}\} + \frac{1}{2}\{ \ket{1}\otimes (I-U)\ket{\Psi}\}
\end{align}
Let the random variable $X$ take the values $X=\pm 1$ when the auxiliary qubit is measured in state $\ket{0},\ket{1}$ respectively. The expected value of $X$ is then
\begin{align}
    \mathbb{E}(X) &= (+1)(Pr(X=+1)) + (-1)(Pr(X=-1))\\
    &= \frac{1}{4}\bra{\Psi} (I+U^\dagger)(I+U) - (I-U^\dagger)(I-U)\ket{\Psi}\\ &= \Re \bra{\Psi}U\ket{\Psi}
\end{align}
and the variance of $X$ is
\begin{align}
    Var(X) &= \mathbb{E}(X^2) - \mathbb{E}(X)^2\\
    &= [(+1)^2Pr(X=+1) + (-1)^2Pr(X=-1)] - [\Re\bra{\Psi}U\ket{\Psi}]^2\\
    &= 1-[\Re\bra{\Psi}U\ket{\Psi}]^2
\end{align}
\tocless\subsection{Measurement with Block Encoding Method}
\label{app:VarianceBlock}
 We discuss the measurement of the expectation $\langle A^\dagger O A\rangle$ and estimation of the variance. We are given $U_A$, a $(1,\ell,0)$ block encoding of $A$, and $U_O$, a $(1,q,0)$ block encoding of the observable $O$ such that $\|O\|\le 1$. We discuss in the main text that $A\ket{\psi}/\sqrt{p}$ is implemented with probability $p$. Let this process be denoted by the Bernoulli random variable $Y=1,0$, such that $Pr(Y=1) = p$ for successfully implementing the state. Once we know the state is successfully implemented by measuring the auxiliary qubit $\ket{0}^{\otimes \ell}$, we can take $\ket{\Psi}=\ket{0}^{\otimes n}A\ket{\psi}/\sqrt{p}$ and perform a Hadamard test with the unitary $U=U_O$ to obtain $\langle A^\dagger O A \rangle/p$. Let $Z$ denote the random variable taking values $Z=\pm 1$ when the auxiliary qubit is measured in state $\ket{0},\ket{1}$ respectively in the Hadamard test. Therefore, $Pr(Z=1|Y=1) = \| (I+U)\ket{\Psi}\|^2/4$ and $Pr(Z=-1|Y=1) = \| (I-U)\ket{\Psi}\|^2/4$. We define the random variable $X = YZ$, where it can be shown that $\mathbb{E}(X) = \langle A^\dagger O A\rangle$ and $Var(X)=p-\langle A^\dagger O A \rangle^2$. Given $N$ shots of the input state $\ket{\psi}$, we can build an estimator of the expectation value of $X$: $\widehat{\langle X \rangle} = \frac{1}{N} \sum_{i=1}^N Y_i Z_i$, where $Y_i,Z_i$ are the random variables associated with the $i^\text{th}$ measurement of the auxiliary qubit $\ket{0}^{\otimes \ell}$ to implement the state successfully and the auxiliary qubit in the Hadamard test respectively. Therefore, 
\begin{align}
    \mathbb{E}(\widehat{\langle X \rangle}) &= \bra{\psi}A^\dagger O A \ket{\psi}, & Var_D(\widehat{\langle X \rangle}) &= \frac{1}{N} \bigg(p-\bra{\psi}A^\dagger O A \ket{\psi}^2 \bigg)\;. 
\end{align}
For each measurement in the worst cast we implement both block encodings $U_A$ requiring at most $\mathcal{O}(L^2d^2)$ and a controlled $U_O$ needing at most $\mathcal{O}(Q^2d^2)$, where $L=2^\ell$ and $Q=2^q$. When $\ell=1,q=1$ such as in the Sz.-Nagy encoding, the measurement requires $\mathcal{O}(d^2)$ one and two-qubit gates in the worst case.
 
\tocless\subsection{Measurement using the Two-Unitary Decomposition Algorithm}
\label{app:VarianceTUD}
Alternatively, we can estimate $\bra{\psi} A^\dagger O A \ket{\psi}$ via the TUD algorithm by measuring the three expectation values below, 
\begin{align}
   \langle  A^\dagger O A \rangle = \frac{1}{4}\{    \langle  U_{1}^\dagger O   U_{1} \rangle +   \langle  U_{2}^\dagger O  U_{2} \rangle +    2\Re \langle  U_{1}^\dagger O  U_{2} \rangle\}
    \;.
\end{align}
Since we are interested in comparing the variances of both methods, we removed the $\epsilon$-approximations of the unitaries which has a negligible effect on this calculation. Each of the terms above are measured individually using three Hadamard tests and are added together to obtain $\langle A^\dagger O A\rangle$. Defining three random variables $X_1,X_2, X_3 \in \{ +1,-1\}$ that denote the auxiliary qubit being measured as $\ket{0}$ or $\ket{1}$ for each Hadamard test such that 
\begin{align}
    \mathbb{E}(X_1) &= \langle U_1^\dagger O U_1 \rangle, & \mathbb{E}(X_2) &= \langle U_2^\dagger O U_2 \rangle, & \mathbb{E}(X_3) &= \Re \langle U_1^\dagger O U_2 \rangle \;.
\end{align}
Let the random variable $X$ be defined such that 
\begin{align}
    X = \frac{1}{4}\{X_1 + X_2 +2 X_3 \}\;, \label{eq:RandomExpect}
\end{align}
where we see that $\mathbb{E}(X)=\langle A^\dagger O A \rangle$. Given $N^\prime$ shots of initial state $\ket{\psi}$, we define the estimators 
\begin{align}
\widehat{\langle X_1 \rangle} &= \frac{4}{N^\prime} \sum_{i=1}^{N^\prime/4} X_{1i},& \widehat{\langle X_2 \rangle} &= \frac{4}{N^\prime} \sum_{i=1}^{N^\prime/4} X_{2i}, & \widehat{\langle X_3 \rangle} &= \frac{2}{N^\prime} \sum_{i=1}^{N^\prime/2} X_{3i}\;,
\end{align}
where $i$ denotes the $i^{\text{th}}$ measurement. Here $X_1, X_2$ are measured $N^\prime/4$ times and $X_3$ is measured $N^\prime/2$ times which is in proportion to their respective weights in Eq.~\eqref{eq:RandomExpect}.
The variance of the estimator $\hat X$ is 
\begin{align}
    Var_{TUD}(\widehat{\langle X \rangle}) &= \frac{1}{N^\prime} \bigg(1-\frac{1}{16}(4\langle U_1^\dagger O U_1\rangle^2 + 4 \langle U_2^\dagger O U_2 \rangle^2 + 8 \Re\langle U_1^\dagger O U_2\rangle^2 )\bigg)
\end{align}

\section{Quantum Singular Value Transformation}\label{app:QSVT}
In this section, we briefly summarize the quantum singular value transform and provide details regarding our specific usage of it.  Readers seeking more information should consult the references cited herein.

The quantum singular value transform (QSVT) \cite{gilyen2019quantum} is an algorithm that allows one to apply a polynomial transformation to the singular values of an arbitrary matrix which is block-encoded inside of a larger unitary.  The intuition for the structure of this algorithm lies with earlier work on quantum signal processing (QSP) \cite{low2016} which studied what kinds of unitary transformations are achievable by alternating between a ``signal'' unitary (which encodes some kind of accessible quantum information) and a single-qubit phase gate with a varying rotation angle (which provides a second control axis).  A rich set of polynomial transformations of the information encoded by the signal unitary was obtained, and this behavior was later extended beyond the single-qubit case, where it was shown that the entire spectrum of a block-encoded Hermitian matrix $H$ could be processed similarly \cite{low2019hamiltonian}.  To achieve this, one first doubles the dimension of the Hilbert space by appending an auxiliary qubit, and then transforms the block-encoding into an iterate that takes the form of a product of two reflection operators.  By Jordan's Lemma, each 1-dimensional eigenspace of $H$ then defines an $SU(2)$ subspace (hence the name ``qubitization'') in which one can generate coherent rotations without mixing into any of the other eigen(sub)spaces.  With additional access to controlled-phase gates, one can achieve the same kind of ``two-axis'' control within each eigen(sub)space, and thus effect the same polynomial transformations of quantum signal processing on each eigenvalue, and therefore the matrix itself, in parallel.  Lastly, these results were further generalized \cite{gilyen2019quantum} to arbitrary (even rectangular) matrices, where by carefully considering the left and right singular vector spaces, the same behavior can be derived with even fewer assumptions.  The importance of these results is well-motivated by recent observations that many quantum algorithms with no overt similarities and widely differing use-cases can be generalized as instances of QSVT \cite{gilyen2019quantum,martyn2021}.

Let us begin by assuming that we have a matrix with SVD form $A = W\Sigma V = \sum_i \sigma_i |w_i\rangle\langle v_i|$ which is block-encoded within a larger unitary $U$, given by $A = \tilde{\Pi} U \Pi$.  The projector $\Pi$ projects into the right singular vector space, i.e. $\text{img}(\Pi) = \sum_i |v_i\rangle\langle v_i|$, and $\text{img}(\tilde{\Pi}) = \sum_i |w_i\rangle\langle w_i|$ thus projects into the left singular vector space.  These projectors then locate $A$ within $U$, and with access to controlled versions of $\Pi$ and $\tilde{\Pi}$ (i.e. rotations conditioned on lying in the span of $\Pi$ or $\tilde{\Pi}$), one can interleave them with $U$ to produce the desired polynomial transformation: $P(A) = \sum_i P(\sigma_i) |w_i\rangle\langle v_i|$.  When restricted to the appropriate invariant $SU(2)$ subspace, the actions of $U$ and the controlled rotations have the form
\begin{align}\label{eq:restricted}
    [U]_{\mathcal{B}_i} = \begin{bmatrix}
        \sigma_i & \sqrt{1-\sigma_i^2} \\
        \sqrt{1-\sigma_i^2} & -\sigma_i
    \end{bmatrix} \qquad
    [e^{i \phi (2\Pi - I)}]_{\mathcal{B}_i} = 
    \begin{bmatrix}
        e^{i \phi} & 0 \\
        0 & e^{-i \phi}
    \end{bmatrix}
\end{align}
where $\mathcal{B}_i$ denotes the $SU(2)$ subspace of the $i$-th singular value.  The operator $[U]_{\mathcal{B}_i}$ is often referred to as the ``iterate'' or the signal operator, and the operator $[e^{i \phi (2\Pi - I)}]_{\mathcal{B}_i}$ is sometimes called the signal processing operator or $S(\phi)$ \cite{martyn2021}.  The actions of $U^\dagger$ and $e^{i \phi (2\tilde{\Pi} - I)}$ are identical but with swapped input/output spaces, with respect to the above operators.  Referring to Definition 8 of \cite{gilyen2019quantum}, we then denote an alternating phase modulation sequence $U_{\Phi}^{(APM)}$ by
\begin{align}
    U_{\Phi}^{(APM)} = 
    \begin{cases}
        U e^{i \phi_n (2\Pi - I)} \dots U^\dagger e^{i \phi_2 (2\tilde{\Pi} - I)} U e^{i \phi_1 (2\Pi - I)} &\quad\text{if $n$ is odd} \\
        U^\dagger e^{i \phi_n (2\tilde{\Pi} - I)} \dots U^\dagger e^{i \phi_2 (2\tilde{\Pi} - I)} U e^{i \phi_1 (2\Pi - I)} &\quad\text{if $n$ is even}
    \end{cases}
\end{align}
The set of angles $\{ \phi_n \}$ determines the resulting degree-$n$ polynomial, and can be efficiently computed \cite{haah2019product,chao2020finding,dong2021}.  We see that the degree of the polynomial determines the number of alternations between $U$ and $U^\dagger$, and thus also determines the final output space (left if odd, right if even).  Once this sequence is constructed, we need only to project into the required block to obtain the transformed operator $P(A)$, where $P$ denotes the desired polynomial (Theorem 10 \cite{gilyen2019quantum}):
\begin{align}
    P(\tilde{\Pi} U \Pi) = 
    \begin{cases}
        \tilde{\Pi} U_{\Phi}^{(APM)} \Pi &\quad\text{if $n$ is odd} \\
        \Pi U_{\Phi}^{(APM)} \Pi &\quad\text{if $n$ is even}
    \end{cases}
\end{align}
The action of the polynomial needs only be correct over the domain $x \in [-1,1]$ due to the fact that $\|A\| < 1$ for $A$ to be properly embedded in $U$, and moreover that singular values are always bounded between 0 and 1.  Because the polynomial produced by QSVT is in general complex ($P \in \mathbb{C}[x]$), it is important to know how to implement real polynomials ($P \in \mathbb{R}[x]$).  This can be achieved by noting that the angle sequence $\{ -\phi_n \}$ produces the conjugate polynomial $P^*$.  One then only needs to perform an LCU-style addition of $U_{\Phi}^{(APM)}$ and $U_{-\Phi}^{(APM)}$ to produce a unitary that block-encodes $\Re(P)$, because the imaginary components will cancel out \cite{gilyen2019quantum}:
\begin{align}\label{eq:qsp_real}
    P_{real}(\tilde{\Pi} U \Pi) = 
    \begin{cases}
        (\langle +| \otimes \tilde{\Pi} )(|0\rangle\langle 0| \otimes U_{\Phi}^{(APM)} + |1\rangle\langle 1| \otimes U_{-\Phi}^{(APM)})(|+\rangle \otimes \Pi) &\quad\text{if $n$ is odd} \\
        (\langle +| \otimes \Pi )(|0\rangle\langle 0| \otimes U_{\Phi}^{(APM)} + |1\rangle\langle 1| \otimes U_{-\Phi}^{(APM)})(|+\rangle \otimes \Pi) &\quad\text{if $n$ is even}
    \end{cases}
\end{align}
The same result can be achieved without the extra LCU qubit by the observation in Appendix B of \cite{dong2021}, where one can simply use a $|+\rangle$ state to control the QSVT auxiliary qubit, and then measure in the $x$-basis to add the two cases.

In general, there exist several conventions in the literature for the QSVT formalism, due to flexibility in how one chooses to define the unitaries used to generate the polynomials.  In \cite{martyn2021}, the authors define four main ingredients, already alluded to above: a signal operator $W$ (the block-encoding), a set of phase angles $\{ \phi_n \}$, a signal \textit{processing} operator $S(\phi)$ (the projector-controlled phase gates), and a basis $M$ in which we measure the QSVT auxiliary qubit to obtain the encoded output.  We can then use $(W,S,M)$ to denote a particular QSP convention. The form of $W$ used in Eq.~\eqref{eq:restricted} is known as the $R$-iterate (for ``reflection''), and results from the particular relations between the left and right spaces proved by \cite{gilyen2019quantum}.  The $R$-iterate can be converted to other conventions, such as $W_x$ or $W_z$, by appropriate single-qubit rotations in the invariant subspace(s).  The equivalence between different QSP conventions, such as $(W_z,S_x,\langle 0| \cdot |0\rangle)$ and $(W_x,S_z,\langle +| \cdot |+\rangle)$, stems from the choice of using either the Laurent or Chebyshev bases for representing polynomials.

\tocless\subsection{QSVT for $f(A)$}\label{app:QSVT_sqrt}

In this section, we provide details regarding our specific implementation of the function $\sqrt{1-x^2}$ using open-source Python repository \texttt{pyqsp} \cite{pyqsp}.  We choose the $(R,S_z,\langle 0| \cdot |0\rangle)$ convention, due to the fact that we presume access to block encodings of the more general format
\begin{align}
    U = \begin{bmatrix}
        \quad A  & \quad \cdot \quad \\
        \quad \cdot & \cdot
    \end{bmatrix}
\end{align}
where the $R$-iterate form implicitly holds by the results of \cite{gilyen2019quantum}.  If one assumes direct access to block-encodings of the Sz.-Nagy form, one immediately recovers the $R$-iterate in the computational basis and can take advantage of measuring in the $\langle + | \cdot | + \rangle$ convention, which is more convenient for representing real polynomials.  In fact, such an assumption immediately gives us our results without needing QSVT at all (Sec.~\ref{sec:assuming_sznagy}).  Other oracles for the block-encoding, such as LCU, require the $\langle 0 | \cdot | 0 \rangle$ convention, which results in complex polynomial outputs that must have their imaginary components removed in the manner of Eq.~\eqref{eq:qsp_real} if one desires a real output.  In particular, this convention is much less convenient because the polynomial outputs must obey two additional constraints \cite{martyn2021, gilyen2019quantum}
\begin{align}
    |P(\pm 1)| &= 1 \\
    P(ix)P^*(ix) &\geq 1 
\end{align}
where $P(\cdot)$ is the implemented polynomial.  Although these constraints result in higher approximation error for a given degree, we choose to use this convention because it is compatible with more realistic oracle models such as LCU.  

\begin{figure}
\centering
\begin{tikzpicture}
\node[scale=0.74] {
\begin{quantikz}[row sep=0.22cm, column sep=0.27cm]
    \ket{0} & \gate{H} & \targ{} & \gate{e^{i\phi_n \sigma_z}} & \targ{} & \qw & \targ{} & \gate{e^{i\phi_{n-1} \sigma_z}} & \targ{} & \qw & \qw & \ \ldots\ \qw & \targ{} & \gate{e^{i\phi_1 \sigma_z}} & \targ{} & \qw & \targ{} & \gate{e^{i\phi_0 \sigma_z}} & \targ{} & \gate{H} & \qw \\
    \ket{0} & \qwbundle{\ell} & \octrl{-1} & \qw & \octrl{-1} & \gate[2]{U_A} & \octrl{-1} & \qw & \octrl{-1} & \gate[2]{U_A^\dagger} & \qw & \ \ldots\ \qw & \octrl{-1} & \qw & \octrl{-1} & \gate[2]{U_A} & \octrl{-1} & \qw & \octrl{-1} & \qw & \qw \\
    \ket{\psi} & \qw & \qw & \qw & \qw & \qw & \qw & \qw & \qw & \qw & \qw & \ \ldots\ \qw & \qw & \qw & \qw & \qw & \qw & \qw & \qw & \qw & \qw &
\end{quantikz}
};
\end{tikzpicture}
\caption{QSVT circuit used for applying real and odd polynomial transformation to the singular values of an arbitrary matrix $A$, in the convention used by \texttt{pyqsp}.  If one desires an even polynomial, the final application of the block-encoding will be $U^\dagger$ instead of $U$.}
\label{fig:qsvt_real_odd_circuit}
\end{figure}
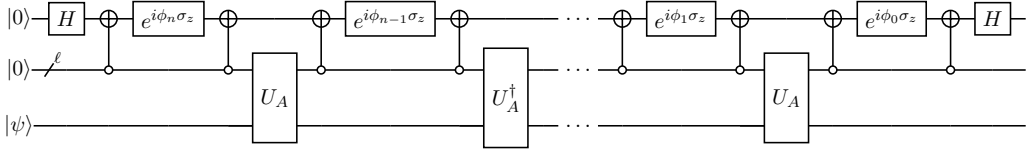

The angles output by \texttt{pyqsp} are in the $W_x$ form, and must be converted to the $R$-form by the following map \cite{martyn2021}: 
\begin{align}
    \phi_i \rightarrow
    \begin{cases}
        \phi_i + \frac{\pi}{4}(2L - 1) & \text{if $i=0$}\\
        \phi_i - \frac{\pi}{4} & \text{if $i=L$}\\
        \phi_i - \frac{\pi}{2} & \text{else}\\
    \end{cases}
\end{align}
The final circuit for the QSVT is shown in Fig.~\ref{fig:qsvt_real_odd_circuit}.  As described in the main text, we must start in the right space and end in the left space to obtain the desired map $\sum_i \sqrt{1-\sigma_i^2} |w_i\rangle\langle v_i|$.  This necessitates either an odd polynomial approximation to the even function $\sqrt{1-x^2}$, or a decomposition of $A$ into Hermitian and anti-Hermitian components so that one resolves the input/output space issue.  {For reproducibility, the 51 angles obtained via \texttt{pyqsp} for the QSVT of Fig.~\ref{fig:odd_poly_51} (\textit{before} using the above conversion) are}
\begin{alltt}
{
    [0.5945105   0.60960954  0.7298763   2.8723814   1.550056    1.2478894
    2.861402    0.76238096  2.014291    1.1355253   2.2245793   3.4903798
    3.044659    0.3553313   3.0701022   0.43656695  0.7831127   2.5623524
    2.491012    0.32272875  3.107868    1.2339981   1.6407303   0.9969318
    1.0959318   2.647887    2.8997867   2.795346    0.04209794  2.7231102
    1.8544827   1.3496944   2.506372    1.5447122   1.5370306  -0.6786906
    0.74581164  0.21720853  2.302732    0.7340103   0.32469824  3.3565025
    2.1487994   2.6228967   0.3943159   0.22044371 -0.06635685  0.17362866
    3.080445    0.7867449   1.2483609   0.24187739]
   }
\end{alltt}
{and the 30 angles for the QSVT of Fig.~\ref{fig:even_poly_30} are}
\begin{alltt}{
    [2.879597    3.2655551   3.1311133   0.20452265  3.5637999   3.1725397
    2.182688    0.55169135  2.6656814   3.470747    3.7887444   0.34566507
    -0.42822552  3.759092    0.41655388 -0.68946004  2.8910365   0.09444891
    2.773645    0.6924566   2.0532732   2.8459213   3.2667234   0.01168075
    3.6077757   3.4687052  -0.6682357  -0.05417808 -0.09065018  3.120173
    1.693089]
    }
\end{alltt}

\tocless\subsection{Complexity for Scaling in Two-Unitary Decomposition}
\label{app:ComplexityTUD}
The complexity for step $1$ in the two-unitary decomposition algorithm that involves the use of QSVT to create the polynomial approximation to the matrix-function $f(A)=\sum_i\text{sign}(\sigma_i)\sqrt{1-\sigma_i^2}\ket{w_i}\bra{v_i}$ depends on the number of oracle calls to the block encoding oracle $U_A$. Since each call to the $U_A$ increases the degree of the polynomial by one, the complexity  by finding the degree $n$ of the $\epsilon$ approximating scalar polynomial $\tilde f(x)$ for the scalar function $f(x)=\text{sign}(x)\sqrt{1-x^2}$. 

Let $f_1(x)=\text{sign}(x)$ and $f_2(x)=\sqrt{1-x^2}$, such that $f(x)=f_1(x)f_2(x)$. Then, Corollary 6 of the Ref.~\cite{low2017hamiltonian} or Ref.~\cite{gilyen2019quantum,martyn2021} show that 
\begin{align}
   \max_{x\in [-1,-\delta_1/2]\cup[\delta_1/2,1]}| f_1(x)-\tilde f_1(x)| \leq \epsilon \;,
\end{align}
where $\tilde f_1(x)$ is a polynomial of degree $n_1=\mathcal{O}(1/\delta_1\log(1/\epsilon))$. We can obtain the condition for the approximation of the function $f_2(x)$ by a $n_2$-degree polynomial $\tilde f_2(x)$ over the domain $x \in [-1+\delta_2,1-\delta_2]$ such that
\begin{align}
    |f_2(x)-\tilde f_2(x)| \leq \sum_{i=n_2}^\infty |\gamma_i x^i| \leq (1-\delta_2)^{n_2} B \leq e^{-n_2\delta} B \leq \epsilon \;.
\end{align}
where $\gamma_i$ are the taylor expansion coefficients and $B=\sum_{i=0}^\infty|\gamma_i|$~\cite{subramanian2019implementing}. Then $n_2 \geq (1/\delta_2\log(B/\epsilon))$~\footnote{A better bound may be obtained by expanding in Chebyshev polynomials.}, since we are concerned with the order of magnitude estimate in terms of $\delta_2$ and $\epsilon$, $n_2 = \mathcal{O}(1/\delta_2)\log(1/\epsilon)$. Let the approximating polynomial of the function $f(x)=f_1(x)f_2(x)$ be given by $\tilde f(x)=\tilde f_1(x) \tilde f_2(x)$, then 
\begin{align}
|f(x)-\tilde f(x)| &= |f_1(x)f_2(x)-\tilde f_1(x)\tilde f_2(x)|\\
&= |\tilde f_1(x) (f_2(x)-\tilde f_2(x)) + \tilde f_2(x) (f_1(x)-\tilde f_1(x))\nonumber \\ &+ (f_1(x)-\tilde f_1(x))(f_2(x)-\tilde f_2(x)) |\\
&\leq 2\epsilon \;,
\end{align}
where we used $|\tilde f_1(x)|,|\tilde f_2(x)| \leq 1$ and neglected the $\mathcal{O}(\epsilon^2)$ error. The degree of approximating polynomial $\tilde f(x)$ is therefore $n = n_1 + n_2 = \mathcal{O}(1/\delta\log(1/\epsilon))$, where we chose $\delta = \min\{\delta_1,\delta_2\}$ such that $x\in [-1+\delta,-\delta]\cup[\delta,1-\delta]$.

\tocless\subsection{Complexity for Scaling in Four Unitary Decomposition}
\label{app:FUDComplexity}
We chose the even function $f(x)=\sqrt{1-x^2}$ to be approximated by an even polynomial to avoid a discontinuity at 0 in the error profile.
The degree $n$ of approximating polynomial can be found as 
\begin{align}
    | f(x)-\tilde f(x)| \leq \sum_{i=n}^{\infty} |\gamma_i \alpha^i x^i/\alpha^i| \leq {e^{-n\delta}}/{\alpha^n} B_\alpha \;,
\end{align}
where we used that $x/\alpha \in [(-1+\delta)/\alpha,(1-\delta)/\alpha]$ and $B_{\alpha} \leq \sum_{i=n}^\infty |\gamma_i \alpha^i|$ is an upper bound on the weighted Taylor series coefficients. We find that $n \geq {1}/(\log\alpha + \delta) \log (B_\alpha/\epsilon)$ for the $\epsilon$ approximation to hold within the domain of $x/\alpha$.
We see that  when we do not know $\delta$ or $\delta=0$, we can chose $n \geq 1/\log(\alpha)\log(B_{\alpha}/\epsilon)\geq{1}/(\log\alpha + \delta) \log (B_\alpha/\epsilon)$ removing the dependence on $\delta$.

However, if we have a matrix that is sufficiently sparse implying only few non-zero eigenvalues, the compression of the eigenvalues will not affect the distribution of eigenvalues much as it is not dense. For this special case, $H/\alpha = H^\prime$ where $H^\prime $ is another sparse matrix such that its eigenvalues $\lambda^\prime_i \in [-1+\delta^\prime,1-\delta^\prime]$ where $\delta^\prime = 1-1/\alpha$. The complexity in this case is then $\mathcal{O}(1/\delta^\prime \log(1/\epsilon))$. 
The number of oracle calls $n$ decreases with increasing $\alpha$ for the $\epsilon$ approximation valid over a smaller domain. However this comes at the cost of the approximation error, as when we scale back we will we have to use $\alpha \tilde U^\prime, \alpha \tilde U^{\prime\dagger}$ in calculation involving $H$ and therefore the error will scale with $\alpha$.

\tocless\subsection{Intuition for Two-Unitary Decomposition}\label{app:intuition}
Rather than appeal directly to SVD to derive the two-unitary decomposition, one can motivate the approach given that we have QSVT in mind.  Since we prefer to work with unitaries, one might ask what kind of processing is needed to produce a unitary from $A$, using only basic block-matrix arithmetic (such as what is defined in Section 5 of \cite{gilyen2019quantum}).  Multiplication of $A$ by some matrix can indeed produce a unitary via the polar decomposition of $A$:
\begin{align}
    A &= UP \rightarrow A^\dagger A = P^\dagger U^\dagger U P \rightarrow P = \sqrt{A^\dagger A}
\end{align}
If $A$ is invertible and if $P$ is the positive-definite square root of $A$, then we can obtain a unitary $U = A P^{-1} = A (\sqrt{A^\dagger A})^{-1}$.  Performing matrix inversion is costly through QSVT, so we may consider trying addition rather than multiplication:
\begin{align}
    U &= A + iM\\
    U^\dagger U &= (A^\dagger - iM^\dagger)(A + iM)\\
    &= A^\dagger A + iA^\dagger M - iM^\dagger A + M^\dagger M
\end{align}
Given that we must always process a block-encoding of $A$, we may assume that $M$ is some function of $A$, i.e. that $M = f(A) = W f(\Sigma) V^\dagger$.  Thus we may write
\begin{align}
    A^\dagger M - M^\dagger A &= V \Sigma W^\dagger W f(\Sigma) V^\dagger - V f(\Sigma) W^\dagger W \Sigma V^\dagger \\
    &= V(\Sigma f(\Sigma) - f(\Sigma) \Sigma)V^\dagger = 0
\end{align}
which follows from the fact that $\Sigma$ is diagonal and thus commutes with $f(\Sigma)$.  We are left with
\begin{align}
    I &= A^\dagger A + M^\dagger M\\
    &= V \Sigma^2 V^\dagger + V f(\Sigma)^2 V^\dagger\\
    f(\Sigma) &= \sqrt{I - \Sigma^2}
\end{align}
which proves that $M = \sqrt{I - A^2}$ is sufficient to obtain a unitary, and avoids the use of matrix inversion.

\end{document}